\documentclass{article}
\usepackage{amssymb}

\usepackage{graphicx}
\usepackage{amsmath}

\input{tcilatex}

\begin{document}

\underline{{\Huge On the AAGL Protocol}}\medskip

M. M. Chowdhury\medskip

\underline{1. Abstract}\medskip

Recently the AAGL (Anshel-Anshel-Goldfeld-Lemieux) has been proposed which
can be used for RFID tags. We give\ algorithms for the problem (we call the
MSCSPv) on which the security of the AAGL protocol is based upon. Hence we
give various attacks for general parameters on the recent AAGL protocol
proposed. One of our attack is a deterministic algorithm which has space
complexity and time complexity both at least exponential in the worst case.
In a better case using a probabilistic algorithm the time complexity can be $%
O(|XSS(u_{i}^{\prime })|^{\lambda _{5}}n^{1+\epsilon })$ and the space
complexity can be $O(|XSS(u_{i}^{\prime })|^{\lambda _{6}})$, where the
element $u_{i}^{\prime }$ is part of a public key,$\ n$ is the index of
braid group, $XSS$ is a summit type set and $\epsilon $ is a constant in a
limit. The above shows the AAGL protocol is potentially not significantly
more secure as using key agreement protocols based on the conjugacy problem
such as the AAG (Anshel-Anshel-Goldfeld) protocol because both protocols can
be broken with complexity which do not significantly differ. We think our
attacks can be improved.\medskip

\underline{2. Introduction}\medskip

\ Recently the AAGL (Anshel-Anshel-Goldfeld-Lemieux) key agreement protocol
using braid groups has been proposed [1] an application of the AAGL protocol
is for RFID\ tags [1]. There is an instantiation of the AAGL protocol in [1]
where the AAGL protocol uses braid groups, in all of this paper we refer to
the AAGL protocol when it uses braid groups. In this note we give an attack
which can show the security of the protocol is based on the multiple
simultaneous conjugacy search problem (see definition 1 below). We think our
attack can be improved. Note once $z$ is recovered with our attack then
agreed upon key can be computed with the linear algebraic attack given in
[1]. All our algorithms can work in groups that are not the braid
group.\medskip

\underline{2.1 Hard Problems in Non-abelian Groups}. \medskip

Definition-The MSCSP (multiple simultaneous conjugacy search problem) [4] is
find elements $g\in G$ such that $y_{i}=gx_{i}g^{-1}$, given the publicly
known information: $G$ is a group, $x_{i},y_{i}\in G$ with $%
x_{i},y_{i}=ax_{i}a^{-1}$, $1\leq i\leq u,$ with the secret element $a\in G.$

Notation-We refer to an example of the MSCSP as $%
((x_{1},x_{2},...,x_{u}),(y_{1},y_{2},...,y_{u}))$ with solution $%
(g,g^{-1}).\medskip $

Definition-\ Consider the following variant of the MSCSP. If $%
(x_{1},x_{2},...,x_{u})$ is unknown in the MSCSP $%
((x_{1},x_{2},...,x_{u}),(y_{1},y_{2},...,y_{u}))$ and we are then to find
the elements $g$. We refer to the above variant of the MSCSP as the MSCSPv
(MSCSP-variant).

Notation-We refer an example of the MSCSPv as $%
((x_{1},x_{2},...,x_{u}),(y_{1},y_{2},...,y_{u}))$ with solution $%
(g,g^{-1}).\medskip $

Definition-The CSP [4] can be defined as the MSCSP with $u=1$.

Notation-We refer to an example of the CSP as $(x,y)$ with solution $%
(g,g^{-1})$.

Notation-In this paper $XSS$ refers a set that potentially contains one or
more solutions for the MSCSP so $XSS$ can refer to a summit type set such as 
$SSS$.

The security of the AAGL protocol is based on the MSCSPv-this is shown
below. Our attack is an algorithm to solve the MSCSPv. The main purpose of
this paper is using an algorithm of deterministic factorial time and space
complexity or a probabilistic algorithm with time complexity $%
O(E(u_{i}^{\prime })n^{2})$ and space complexity $O(E(u_{i}^{\prime }))$ ($%
O(E(u_{i}^{\prime }))$ is of at least exponential complexity in the braid
index $n$ and its word length $W$ of some braid, so $E(u_{i}^{\prime })$
grows at most, like a power of the factorial of $n$, $O(n!^{W})$, in this
paper we refer to $O(n!^{W})$ as the abbreviation factorial complexity), it
is then to shown the security of the AAGL protocol is equivalent to solving
the MSCSP (and hence the CSP) instead of the MSCSPv.

Our result is better than all previous results in the connection: it works
for general parameters, that there is only a brute force algorithm (which
has factorial complexity) to solve the MSCSPv and our algorithms are better
than the brute force algorithm. The above factorial complexity algorithm of
ours may use any factorial time algorithm for the CSP, note the best
algorithm to solve the CSP in general has in the worst case factorial
running time. Hence this means the AAGL protocol is no more secure than
using the AAG protocol [2] in the connections:

$\bullet $ We show they are both protocols are based on the MSCSP (so the
AAGL protocol is not strictly based on the MSCSPv as implied in [1]).

$\bullet $ They can both be broken using related deterministic algorithms of
factorial complexity that solve the MSCSP.

$\bullet $ There are related probabilistic algorithms (including our
probabilistic algorithm 4) that may break both protocols depending on the
parameters used.\medskip

\underline{3. AAGL Key Agreement Protocol}\medskip

In the recent [1] AAGL\ propose a key agreement protocol it differs mainly
from the seminal AAG (Anshel-Anshel-Goldfeld) algebraic protocol given in
[2] because it is based on the MSCSPv, the AAG protocol is based on a system
of conjugacy equations (the MSCSP) [2]. We do not reproduce all details of
the AAGL\ protocol which can be found in [1] but restrict to the details we
require. Let $B_{n}=\{b_{1},b_{2},...,b_{n-1}\}$ be the Artin representation
of the braid group on $n$ strings. In [1] an example of the protocol is
given using braid groups the security is based on the TTP\ algorithm in [1]
given below. $e$ is the identity element in the braid group.\medskip

\underline{Algorithm 1- TTP Algorithm of [1].}\medskip

1. Choose two secret subset $BL=\{b_{l_{1}},...,b_{l_{\alpha
}}\},BR=\{b_{r_{1}},...,b_{r_{\beta }}\}$ of the set of generators of $B_{n}$
where $|l_{i}-r_{j}|\geq 2$ for all $1\leq i\leq l_{a}$ and $1\leq j\leq
r_{\beta }$.

2. chooses a secret element $z\in B_{n}$.

3. Choose words $\{w_{1},...,w_{\gamma }\}$ of bounded length from $BL$.

4. Choose words $\{v_{1},...,v_{\gamma }\}$ of bounded length from $BR$.

5. For $1\leq i\leq \gamma $

a. calculate the left normal form $zw_{i}z^{-1}$ and reduce the result
modulo the square of the fundamental braid.

b. set $w_{i}^{\prime }$ equal to the sequence of integers that correspond
to the element calculated in a.

c. calculate the left normal form $zv_{i}z^{-1}$ and reduce the result
modulo the square of the fundamental braid.

d. set $v_{i}^{\prime }$ equal to the sequence of integers that correspond
to the element calculated in c.

6. Publish the two sets $\{w_{1}^{\prime },...,w_{\gamma }^{\prime }\}$ and $%
\{v_{1}^{\prime },...,v_{\gamma }^{\prime }\}$.\medskip

The security of the TTP algorithm is based on the MSCSPv with the elements $%
(x_{1},x_{2},...,x_{u})=(w_{1},...,w_{\gamma },v_{1},...,v_{\gamma }),$

$(y_{1},y_{2},...,y_{u})=(w_{1}^{\prime },...,w_{\gamma }^{\prime
},v_{1}^{\prime },...,v_{\gamma }^{\prime })$ and $u=2\gamma $. Assume the
attacker knows this instance of the MSCSPv in Artin representation.\medskip

\underline{3.1 Security of AAGL Protocol is Based on the }

\underline{Multiple Simultaneous Conjugacy Search Problem}\medskip

Notation-$u_{i}^{\prime }\in \{w_{1}^{\prime },...,w_{\gamma }^{\prime
}\}\cup \{v_{1}^{\prime },...,v_{\gamma }^{\prime }\},u_{i}\in
\{w_{1},...,w_{\gamma }\}\cup \{v_{1},...,v_{\gamma }\}$, $u_{i}^{^{\prime
}}\sim u$, for some $1$ $\leq i\leq 2\gamma $.

Notation-$S_{u_{i}^{\prime }}$ is a set that contains elements of the form $%
zkz^{-1}$ when the set is used in a deterministic algorithm. $%
S_{u_{i}^{\prime }}$ contains elements of the form $zkz^{-1}$ with some
probability when the set is used in a probabilistic algorithm. Where $%
zkz^{-1}$are elements in the centraliser of $u_{i}^{\prime }$.

Recall the centraliser of an element is the set of all elements that commute
with it, for the infinite braid group the centraliser of an element will
contain an infinite number of elements hence we approximate the centraliser
with a finite set. Let $\lambda $ be a braid invariant, it is possible (but
unlikely) for two different braids have the same value for $\lambda $ see
[6]. Note there are practical algorithms to compute the braid invariant $%
\lambda $ because the CDP (conjugacy decision problem) is feasible in braid
groups. Good bounds for summit type sets are not known but the SSS certainly
has the upper bound $n!^{q},$where $q=\max_{\forall i}(\min \sup
(u_{i}^{\prime })+\max \inf (u_{i}^{\prime })),$for example see [6], so all
known algorithms for computing SSS are in the worst case have factorial
complexity but it is conjectured that the size $SSS$ is exponential in $n$.
Because we use the upper bound $n!^{q}$ our algorithm is of factorial
complexity. It is known the quantity $\min \sup (u_{i}^{\prime })+\max \inf
(u_{i}^{\prime })$ can be computed in polynomial time and space so $q$ can
be computed in polynomial time and space in $n$ and the length of $%
u_{i}^{\prime }$ (obviously this means $q$ can be computed in factorial
space and time complexity). Refinements of Garside's algorithm for the
CSP/CDP [8] (Garside's is the first algorithmic solution of CDP/CSP) to
solve the CSP such as the solution given in [7] can be used to solve
deterministically the CDP in worst case factorial time and space complexity.
Elements of the form $zkz^{-1}$ (where $k$ commutes with $u_{i}$) may be
found by computing the centraliser of $u_{i}^{\prime }$ (which any attacker
can compute) because $zkz^{-1}zu_{i}z^{-1}=zu_{i}z^{-1}zkz^{-1}\Rightarrow
zkz^{-1}u_{i}^{\prime }=u_{i}^{\prime }zkz^{-1}$ so $zkz^{-1}$ is in $%
S_{u_{i}^{\prime }}$.

\underline{3.1.1 Attack Based on MSCSP}\medskip

The only known attack, given in [1], without side information on the AAGL is
a brute force attack the above linear algebraic attack is given in section 6
of [1]. We give a deterministic algorithm based on computing centralizers to
solve the MSCSPv and our algorithm maybe uses algorithms that compute super
summit sets. Our algorithm has factorial complexity some reasons are because
all known algorithms to compute the centralizer of an element are factorial
complexity (in the worst case)/the best known algorithm to solve the CSP in
general has factorial running time, which means computing centralisers and
solving the CSP can take around the same time. Hence our algorithm is the
best known way to attack the AAGL if suitable parameters can be found/and
potentially its efficiency improved.

Notation-$C$ is an algorithm that computes $S_{u_{i}^{\prime }}$ in
factorial space and time complexity in a worst case.\medskip

\underline{Algorithm 2 General deterministic algorithm for MSCSPv.}\bigskip

1. Compute $S_{u_{i}^{\prime }}$ for $u_{i}^{\prime }=zu_{i}z^{-1}$, $%
S_{u_{i}^{\prime }}$ contains some or all elements of the form $F=zkz^{-1}$
it follows choices for $k$ includes all elements which commute with $u_{i}$
etc.,\thinspace $S_{u_{i}^{\prime }}\subseteq S_{F}.$

2.\ For an element $u_{i}^{\prime }$ in $S_{u_{i}^{\prime }}$ find $k$ then
solve the CSP with ($k,zkz^{-1}$) for $(z,z^{-1}).$ We find $k$ as follows.

2i. Select a function $f_{P}$ which parametrizes in $P$ a finite
approximation to the centralizer $u_{i}^{\prime }$.

2ii. Select a function which parametrizes in $L_{P}$ a finite approximation
to words in $G$. We define by the set $U_{L_{P}}$ as containing all words
defined by $L_{P}$.

3i. Set $L_{P}=L_{0}$. $P=P_{0}.$ $L_{P}$ may depend on $P$. Compute if
necessary $S_{u_{i}^{\prime }}=f_{P}$.

3ii. Update value $P$ as, $P\in p_{v}.$ Initialise $I=I_{0}$.

3iii. Select $S_{u_{i}^{\prime },L_{P}}^{\prime }\subseteq S_{u_{i}^{\prime
}}$. compute if necessary $S_{u_{i}^{\prime }}=f_{P}$. $S_{u_{i}^{\prime
},L_{P}}^{\prime }$ may depend on $L_{P}$. Using a chosen algorithm, find a
(CSP) pair $(b,a)$ such that 
\begin{equation*}
b\sim a,\text{ }a\in S_{u_{i}^{\prime },L_{P}}^{\prime }\text{, }b\in
U_{L_{P}}^{\prime }
\end{equation*}
where $U_{L_{P}}^{\prime }\subseteq U_{L_{P}}$. The pair $(b,a)$ is stored.

3iv. If the values of $P$ have been exhausted from the set $p_{v}$ then goto
step 4.

3v. Update value of $I$ as $I\in i_{v}$, if the values of $I$ have been
exhausted from the set $i_{v}$ then goto step 3ii.

3vi. Update $L_{P}$ as $L_{P}=L_{P,I}$. If the values of $L_{P}$ have not
been exhausted goto step 3iii.

4. Solve the MSCSP for all the pairs $(b,a)$.\medskip\ Terminate algorithm.

We now give an example of algorithm 2 which is also a general algorithm
(note $P$ is redundant in this example for the reason given in the proof
below) of the above algorithm where $G$ may be $B_{n}$ or any Garside
group.\medskip

\underline{Algorithm 3 An example of algorithm 2.}\medskip

1. Compute $S_{u_{i}^{\prime }}$ for $u_{i}^{\prime }=zu_{i}z^{-1}$, $%
S_{u_{i}^{\prime }}$ contains some or all elements of the form $F=zkz^{-1}$
hence for choices of $k$ includes all elements in $BR$ or $BL$ depending if $%
u_{i}^{\prime }=v_{i}^{\prime }$ or $u_{i}^{\prime }=w_{i}^{\prime }$.

A second possible choice is to compute $S_{u_{i}^{\prime }}$ represented by
a generating set of the centraliser of $u_{i}^{\prime }$ such as using the
algorithm in [9].

2.\ Find $k$ then solve the CSP with $(k,zkz^{-1}$) for $(z,z^{-1}).$ We
find $k$ as follows.

2i. Select a function which parametrizes in $P$ a finite approximation to
the centralizer $zu_{i}z^{-1}$. We choose to the function $%
F_{u_{i}^{^{\prime }},P}(P_{0},\alpha $) which computes the set which
contains all braids $F\in S_{F}$ in the centralizer of $u_{i}^{\prime }$
such that $\ \Delta ^{P}\preccurlyeq F\preccurlyeq \Delta ^{P+1}$, for $%
\forall P$, $P_{0}\leq P<\alpha $, which is $\Delta ^{P_{0}}\preccurlyeq
F\preccurlyeq \Delta ^{\alpha }$. $S_{F}$ here contains at least one element
in the centraliser of $u_{i}^{\prime }$ if using $C$ as described in the
proof below.

2ii. We define the set (we construct) $U_{L_{P}}$ as $U_{L_{P}}\subseteq
B_{n}^{+}\backslash e$ to contain all distinct words in Artin of length $%
L_{P}$ with the length is in the number of Artin generators. Or a second
possible choice for $U_{L_{P}}$ may contain some or all of the union of the
centralisers of short words in Artin generators, so for example, to compute $%
k$ is to (where $k$ may be long) choose it from the generating set of the
centraliser of the single Artin generators $\sigma _{i}$ say using the
algorithm in [9], the above approach can be used when $k$ not one Artin
generator.

3i. Set $L_{P}=1.$ $P_{0}=-2g_{z}$. $P=P_{0}$. Let $S_{u_{i}^{\prime
}}=F_{u_{i}^{^{\prime }},P}(P_{0},\alpha )$.

3ii. $P=P+1$. $I=1$.

3iii. We test the relation using an algorithm for the CDP (an alternative
step instead of this step is described in the proof below) 
\begin{equation*}
\lambda (a)=\lambda (b),\text{ }a\in S_{u_{i}^{\prime },L_{P}}^{\prime }%
\text{, }b\in U_{L_{P}}^{\prime }
\end{equation*}
where $S_{u_{i}^{\prime },L_{P}}^{\prime }\subseteq S_{u_{i}^{\prime }}$
,and $U_{L_{P}}^{\prime }\subseteq U_{L_{P}}$, The pair $(b,a)$ are found
with a linear search. If the above relation is true then let $k=b$. The pair 
$(b,a)$ is stored.

3iv. If $P>P_{0}+1$ then goto step 4.

3v. $I=I+1$. $L_{P,I}=I$.

3vi. $L_{P}=L_{P,I}$. If $L_{P}>f(u_{i}^{\prime })$ then goto step 3ii$.$
Where $f(u_{i}^{\prime })$ may depend on $u_{i}^{\prime }$ .

4. Solve the MSCSP for all the pairs $(b,a)$ using a deterministic
algorithm. Terminate algorithm.\medskip

As would be expected, for poorly chosen parameters our algorithm may not be
more efficient than a brute force attack. A potential variant is to check
if\ a short word in length of Artin generators is not conjugated by $z$ an
attacker can compute then length for a given length function the average (or
an upper bound) length of $u_{i}^{\prime }$ and if a word $u_{i}^{\prime }$
is significantly larger than the average length (or an upper bound) of $%
u_{i}^{\prime }$ \ it is considered not a potential value for $a$, so here $%
S_{u_{i}^{\prime },L_{P}}^{\prime }$ depends on $L_{P}$. Other potential
variants (which we discuss below) is$\ $some subset $S_{F\text{ }}$ in the
bounds for $F$ such that $\Delta ^{P_{0}}$ $\preccurlyeq F\preccurlyeq
\Delta ^{\delta }$ (so here $P$ may have a larger range), for some integer $%
\delta $, and not restrict $b$ to positive words. The parameters $P,L_{P}$
control the lengths of $a$ and $b$. We now show for suitable parameters our
attack will terminate with a solution for the MSCSPv used in the AAGL
protocol.\medskip

\underline{Proposition 1}\medskip

Solving the MSCSPv as used in the AAGL protocol is equivalent to solving
MSCSP (which can be shown in deterministic factorial time) using algorithm 3
twice and possibly using algorithm $C$, with the parameters $P_{0}=-2g_{z}$, 
$\alpha =2g_{z}+L_{m}$,$\ f(u_{i}^{\prime })\leq O(\frac{n}{\log (n)})$
where $L_{m}=\max_{\forall P}L_{P}$. As is shown in the proof $L_{m}=1$ is
sufficient. This requires in the worst case, space complexity 
\begin{equation*}
O(c_{1}|SSS(u_{i}^{\prime })|+c_{2}|SSS(b)|+(n-1)^{O(\frac{n}{\log (n)})})
\end{equation*}
and time complexity 
\begin{equation*}
O(|SSS(b)||SSS(u_{i}^{\prime })|(n-1)^{O(\frac{n}{\log (n)})})
\end{equation*}
$.$ Where the element $u_{i}^{\prime }$ is part of the TTP's public key and $%
b\in S_{v_{i}^{\prime }}\cup S_{w_{j}^{\prime }}.$

\underline{Proof}\medskip

We use algorithm that computes summit type sets we call $XSS$ we analyze the
cases for $XSS$, $SSS$ and $C$ both in the proof below.\medskip

$\bullet $ Case using $C$ and $SSS$.\medskip

We use an algorithm $C$ such that it computes $\ \Delta ^{P_{0}}\preccurlyeq
F\preccurlyeq \Delta ^{\alpha }$ for all $P$ used. For $C$ we use an
existing algorithm for the centraliser or the CSP. $\inf_{ss}(u_{i}^{\prime
})$ means elements of the $SS$ which have maximum infinimum of the conjugacy
class of $u_{i}^{\prime }$. For example we use the algorithm to compute the $%
SS$ [8] to solve\ the CSP\ ($u_{i}^{\prime },u_{i}^{\prime }$) then it
follows for $\forall F\in S_{F},$ $\Delta ^{P_{0}}\preccurlyeq F\prec \Delta
^{P_{0}+1}$ where $P_{0}=\inf_{ss}(u_{i}^{\prime })$ and the $SS$ can be
computed in worst case factorial complexity, so $P$ is redundant in this
case, but here $\inf (zkz^{-1})=\inf_{ss}(u_{i}^{\prime })$ must be true for
the algorithm to find a solution. Another choice for $S_{u_{i}^{\prime }}$
involves computing all braids $F,$%
\begin{equation}
\Delta ^{P_{0}}\preccurlyeq F\preccurlyeq \Delta ^{\alpha }  \tag{1}
\end{equation}
then it would follows from the analysis and the bound on the number of
braids in canonical factors and the braid index [3] our algorithm would be
of factorial complexity but this choice of parameters for the algorithm
results in complexity similar to a brute force algorithm. The above method
using 1 can be potentially improved for example if $k$ (described below) is
short length, then depending on the method of rewriting elements,\ $F$ is
expected of short length hence the value of $\alpha $ can be lowered.

In the following analysis we assume the algorithm $C$ is used to compute $%
F\in S_{u_{i}^{\prime }}$, with $F$ in the bounds given by 1, for all $P$
used or the choice for computing $S_{u_{i}^{\prime }}$ above or we assume
our analysis below we use the $SSS$ based algorithm of [7] which has worst
case factorial time and factorial space complexity. If $C$ computes $%
S_{u_{i}^{\prime }}$ contains at least one element of the form $zkz^{-1}$ is
used then our algorithm always terminates with a solution using the bounds
for $P_{0}$,$\alpha $ derived below.

We prove below that computing $U_{L_{P}}$ has worst case factorial time and
factorial space complexity. Note if it is true at least for a class of
braids for a value of $XSS$ that $|XSS|$ is exponential in $n$ in the proof
below then the algorithm works in worst case exponential time (hence the
term $O(\frac{n}{\log (n)})$ appears below).

From [1] any attacker can compute the smallest $g_{z}$ from 
\begin{equation*}
g_{z}\frac{\ln (2n-2)}{\ln (2)}\geq m
\end{equation*}
where $g_{z}$ is the length of $z$ in it Artin generators and we assume the
smallest $g_{z}$ is used. If the above assumption turns out to be false then
the attacker may estimate $g_{z}$ from the elements $u_{i}^{\prime }$, $%
g_{z} $ can be feasibly computed otherwise the public keys are too long in
Artin generators to use.

We use the easy theorem 1.5 given in [7] which is if $B$ is any braid word
represented in $N$ negative Artin generators $P$ positive Artin generators
and then $\Delta ^{N}$ $\preccurlyeq B\preccurlyeq \Delta ^{P}$. Consider
then for any $b\in U_{L_{P}}$, and $z$ then $\Delta ^{-g_{z}}$ $\preccurlyeq
z\preccurlyeq \Delta ^{g_{z}}$, $e$ $\preccurlyeq b\preccurlyeq \Delta
^{L_{p}},$ hence 
\begin{equation*}
\Delta ^{-2g_{z}}\preccurlyeq zbz^{-1}\preccurlyeq \Delta ^{2g_{z}+L_{p}}
\end{equation*}
hence if we let $P_{0}=-2g_{z}$, $\alpha =2g_{z}+L_{p}<2g_{z}+L_{m}$ from
the above bound on $zbz^{-1}$ then it must be true that the centraliser of$\
u_{i}^{\prime }$ contains\ elements of the form $zkz^{-1}$ where $k$ is an
element in $BL$ or $BR$. \ 

Note $|S_{F}|$ by assumption is of factorial complexity in the parameters $%
2g_{z}+L_{p},$ $n$ and contains $zkz^{-1}$ for values of $k.$

This is true because we know from the condition $|l_{i}-r_{j}|\geq 2$ it
follows that $BL$ and $BR$ do not have any generator in common.

Now the attacker runs the algorithm 3 twice in parallel so that $%
u_{i}^{\prime }=v_{i}^{\prime }$ or $u_{i}^{\prime }=w_{j}^{\prime }$ in
each the runs but the attacker may compute the common computations (such as
computing $U_{L_{P}}$) once in each of the runs. Hence for $L=1$ one of the
choices of $\ b$ (selected by the attacker) is one of the Artin generators
of $BR$ must be a correct choice, generally this means for $k$ there are $%
l_{\alpha }+r_{\beta }$ easy to guess when it is of feasibly computable
length from the parameters suggested in [1]. Such easy choices for $k$ exist
because the TTP algorithm specifies the subgroup in terms of single Artin
generators. The attacker using linear search algorithm, through $%
S_{v_{i}^{\prime }}\cup S_{w_{j}^{\prime }}$ for $zkz^{-1}$ and using the
CDP with $b$ a word in a single Artin generator of up length $n$ in step
3iii finds the CSP pair $(b,zbz^{-1})$ to solve which must exist by
construction. An option at step 3iii is to solve the CDP in deterministic
factorial time using the algorithm in [7]. \ 

We estimate $|U_{L_{p}}|$ with the upper bound $S_{n,L_{P}}=(n-1)^{L_{P}}$
hence $S_{n,O(\frac{n}{\log (n)})}$ is at worst an exponential function.
Hence computing all distinct words (which are not\ optimally bounded above
exponentially by $S_{n,L_{P}}$) of length $\theta =O(\frac{n}{\log (n)})$
from $U_{L_{P}}^{\prime }$ will mean the attacker is guaranteed to find all
the words (there are an exponential amount of these and if $\theta =O(\log
(n))$ they are a polynomial long) in $BL,BR$ of length $\theta $ in $%
S_{v_{i}^{\prime }}\cup S_{w_{j}^{\prime }}$.

Hence the attacker can take $1\leq L_{p}\leq O(\frac{n}{\log (n)}),$ as this
keeps the complexity exponential so the attacker selects $L_{P}$ up to $O(%
\frac{n}{\log (n)})$ (actually the attacker can select any $\theta $ such
that the complexity of the algorithm is factorial in the worst case). To get
more conjugacy equations the attacker can try for $b$ all words of length $O(%
\frac{n}{\log (n)})$ but as expected the longer the word length of $b$ is
the less chance (as described below) that a CSP pair will be found but for
short word length of $b$ there is a non-negligible probability that the
attacker can guess a correct $b$. We show below that $1\leq L\leq O(\log n)$
can be chosen.

In the following $\theta =O(\frac{n}{\log (n)})$. Let $\
c_{1},c_{2}...,\lambda _{1},\lambda _{2},...$ $\in \Re $, we assume we may
approximate $S_{u_{i}^{\prime }}$ and centraliser computations by $%
O(|XSS(u_{i}^{\prime })|)$ this assumption is based on the fact that in some
cases, e.g. $XSS=SSS,$ it is known algorithms to computing the centraliser
of $u_{i}^{\prime }$ are proportional in space and time complexity to $%
|XSS(u_{i}^{\prime })|$. $|S_{u_{i}^{\prime }}|$ and hence any centraliser
is at most of space factorial complexity using (2.1) above. $%
SSS(u_{i}^{\prime })$ is the super summit set of the element $u_{i}^{\prime
} $ the size of these sets are not fully known it is known that $|SSS|$ to
be at least exponential in $n$, for a fixed $n$ is proportional to $(n!)^{q}$%
, we write $SSS(u_{i}^{\prime })$ for the maximum size of $SSS(u_{i}^{\prime
}) $ where $q=\max_{\forall i}\min \sup (u_{i}^{\prime })+\max \inf
(u_{i}^{\prime })$. Observe $O((n!)^{q})$ is of smaller order than $%
O(e^{qn\log (n)})$ we use a similar notation for $XSS(u_{i}^{\prime })$.

If we use an algorithm that stores at least all the elements in $%
S_{v_{i}^{\prime }}\cup S_{w_{j}^{\prime }}$ and stores all elements in $%
U_{L_{P}}$, uses an deterministic algorithm to solve both the MSCSP (the
number of equations $\nu $ in the MSCSP may be constant) and CDP that uses
exponential space, then the space complexity of the algorithm is factorial
in the worst case it is 
\begin{eqnarray*}
&&O(c_{1}E(u_{i}^{\prime })+c_{2}E(b)+c_{3}(n-1)^{\theta }) \\
&=&O(c_{4}|SSS(u_{i}^{\prime })|+c_{5}|SSS(b)|+(n-1)^{O(\frac{n}{\log (n)})})
\\
&\thickapprox &O(c_{6}e^{q\log (n)}+c_{7}e^{q_{2}\log (n)}+c_{8}\omega ^{n})
\end{eqnarray*}
where the constant $\omega $ depends on the function used for $O(\frac{n}{%
\log (n)})$.

Note the space and time complexity of solving the MSCSP is proportional to $%
\nu |SSS(b)|$ if the MSCSP solved by intersections of elements of summit
sets.

We write $O(|SSS(b)|)\thickapprox O(E(b))=O(e^{q_{2}n\log (n)})$ as the
maximum size of $SSS(b)$ which is determined by $q_{2}=\max_{b\in
U_{L_{P}}^{\prime }}\min \sup (b)+\max \inf (b),$ so $q_{2}$ is the
canonical length of $b$.

We use this result at step 3iii in the CDP. Computing $S_{v_{i}^{\prime
}}\cup S_{w_{j}^{\prime }}$, the linear searches through $S_{v_{i}^{\prime
}}\cup S_{w_{j}^{\prime }}$ for each element of $U_{L_{P}}$ and solving the
CDP for every potential pair at 3iii, and then solving the resulting MSCSP
means in the worst case the time complexity is factorial and it is 
\begin{eqnarray*}
&&O(c_{1}E(u_{i}^{\prime })+c_{2}E(u_{i}^{\prime })E(b)(n-1)^{\theta }) \\
&\thickapprox &O(E(b)E(u_{i}^{\prime })(n-1)^{O(\frac{n}{\log (n)})}) \\
&=&O(E(b)E(u_{i}^{\prime })\omega ^{n})=O(|SSS(b)||SSS(u_{i}^{\prime
})|\omega ^{n}) \\
&\thickapprox &O((\omega e^{(q+q_{2})\log (n)})^{n})
\end{eqnarray*}
It is understood the constants $c_{1},c_{2}...$ are different from the
constants $c_{1},c_{2}...$ used in the notation of the space complexity and
other complexity computations below. Now consider a variant of the above
attack, which is not use an algorithm for the CDP in step 3iii but instead
solves the CSP with the guess for $b$ with every possible element in $%
S_{u_{i}^{\prime },L_{P}}^{\prime }$ and recovers $z$ and hence the shared
secret using the algorithm in [1], the attacker may test if $z$ is the
correct solution: for example, $z$ is used in an impersonation attack or if
(using an algorithm for the CDP) $z^{-1}u_{i}^{\prime }z\sim u_{i}^{\prime }$%
. For the variant attack above the worst case space and time complexity are
the same.\medskip

$\bullet $XSS Case.\medskip

In the more general case of $XSS$ we can use any algorithm we refer to as $S$
that and outputs for a conjugate pair $(b,a)$ or potential pair $(b,a)$
using the sets $S_{v_{i}^{\prime }},S_{w_{j}^{\prime }},U_{L_{P}}^{\prime }$%
, let $S_{t,u_{i}^{\prime }},S_{s,u_{i}^{\prime }}$ be the time complexity
and space complexity respectively to compute $S_{u_{i}^{\prime }}$, we refer
to the time complexity and space complexity to solve the CDP as $CDP_{t}$, $%
CDP_{s}$, we refer to an algorithm for $CDP$ with input $b,a$ as $CDP(b,a)$
similar notation for $MSCSP.$ Note it is assumed the CDP and MSCSP can be
solved in worst case time complexity and space complexity proportional to $%
|XSS|$, this is true for example when $XSS=SSS$ and the assumption is based
on this example. By a similar argument to the $SSS$ a time complexity bound
in the worst case is, 
\begin{eqnarray*}
&&O(S_{t,v_{i}^{\prime }}+S_{t,w_{j}^{\prime }}+\max_{\forall P\in
p_{v},b\in U_{L_{P}}^{\prime }}CDP_{t}(S(S_{v_{i}^{\prime }}\cup
S_{w_{j}^{\prime }},b))+MSCSP_{t}) \\
&\thickapprox &O(c_{1}|XSS(u_{i}^{\prime })|^{\lambda _{1}}+c_{2}|XSS(b)|)
\end{eqnarray*}
A bound for the space complexity in worst case is 
\begin{eqnarray*}
&&O(S_{s,v_{i}^{\prime }}+S_{s,w_{j}^{\prime }}+U_{s,L_{m}}^{\prime
}+MSCSP_{s}+\max_{\forall P\in p_{v},a\in S_{v_{i}^{\prime }}\cup
S_{w_{j}^{\prime }}}\max_{b\in U_{L_{P}}^{\prime }}CDP_{s}(b,a)) \\
&\thickapprox &O(c_{1}|XSS(u_{i}^{\prime })|^{\lambda
_{2}}+c_{2}|SSS(b)|+c_{3}|U_{L_{P}}^{\prime }|).
\end{eqnarray*}
Now consider a variant of the above attack, which is not use an algorithm
for the CDP in step 3iii but instead solves the CSP with the guess for $b$
with every possible element in $S_{u_{i}^{\prime },L_{P}}^{\prime }$ and
recovers $z$ and hence the shared secret using the algorithm in [1], the
attacker may test if $z$ is the correct solution: for example, $z$ is used
in an impersonation attack or if $z^{-1}u_{i}^{\prime }z\sim u_{i}^{\prime }$%
.

Note it may be that $b\in S_{u_{i}^{\prime },L_{P}}^{\prime }$ (which can be
verified using a polynomial time word algorithm in $B_{n}$), in this case $z$
must be in the centraliser of $b$, call the set of all such stored $b,$ $%
B_{z}$ and so $z$ can be found by testing every element (for the choice of $%
z $) of the centraliser\ of a subset of $B_{z}$ for the correct element.

\ If $S_{u_{i}^{\prime }}$ is computed using the second choice in step 1 and
the corresponding value of $k$ is found using the second choice in step 2ii
then because it is known the centraliser of every element in $B_{n}$ can be
generated by $O(n^{2})$ generators hence the MSCSPv can be solved feasibly
depending on $|SSS|$ [9].\medskip

\underline{Algorithm 4- Probabilistic algorithm for MSCSPv.}\medskip

1. Compute $S_{u_{i}^{\prime }}$ a suitably small $S_{u_{i}^{\prime },s}$ of 
$u_{i}^{\prime }=zu_{i}z^{-1}$ that may contain elements of the form $%
F=zkz^{-1}$ hence for choices of $k$ includes all elements in $BR$ or $BL$
depending if $u_{i}^{\prime }=v_{i}^{\prime }$ or $u_{i}^{\prime
}=w_{i}^{\prime }$. One possible simple choice at this step is to compute $%
S_{u_{i}^{\prime }}$ as randomly chosen elements of the centraliser of $%
u_{i}^{\prime }$.

2.\ Find $k$ then solve the CSP with $(k,zkz^{-1}$) for $(z,z^{-1}).$ We
find $k$ as follows.

2i. Select a function which parametrizes in $P$ a suitably small finite
approximation to the centralizer $zu_{i}z^{-1}.$ We choose the function $%
F_{u_{i}^{^{\prime }},P}(P_{0},\alpha )$ which computes the set which
contains a subset of the braids $F\in S_{F}$ and if possible maybe using
heuristic method(s) gives $F$ where $k$ is short (in a given length
function) with high probability.

2ii. We define the set we can feasibly compute $U_{L_{P}}$ as $%
U_{L_{P}}\subseteq B_{n}$ of short words in length $L_{P}$ for some length
function.

3i. Set $L_{P}=1.$ $P_{0}=-2g_{z}$. $P=P_{0}$. Let $S_{u_{i}^{\prime
}}=F_{u_{i}^{^{\prime }},P}(P_{0},\alpha )$.

3ii. $P=P+1$. $I=1$.

3iii. We test the relation using an efficient algorithm to solve the CDP
such as the one in [6] 
\begin{equation*}
\lambda (a)=\lambda (b),\text{ }a\in S_{u_{i}^{\prime },L_{P}}^{\prime }%
\text{, }b\in U_{L_{P}}^{\prime }
\end{equation*}
where $S_{u_{i}^{\prime },L_{P}}^{\prime }\subseteq S_{u_{i}^{\prime }}$
,and $U_{L_{P}}^{\prime }\subseteq U_{L_{P}}$. If the above relation is true
then let $k=b$. The pair $(b,a)$ is stored. When enough pairs have been
computed goto step 4.

3iv. $I=I+1$.

3v. $L_{P,I}=I$.

3vi. $L_{P}=L_{P,I}$. If $P>P_{0}$ then terminate then goto 4. If $%
L_{P}>f(u_{i}^{\prime })$ then goto step 3ii$.$ Where $f(u_{i}^{\prime })$
may depend on $u_{i}^{\prime }$ we can take it to be on feasibly computable
words up to $O(\frac{n}{\log (n)}).$

4. Solve the MSCSP for all the pairs $(b,a)$ using an algorithm that works
with high probability. Terminate algorithm.\medskip

\underline{Proposition 2}\medskip

Solving the MSCSPv can be done with the probabilistic algorithm 4, in
approximately, time $O(|XSS(u_{i}^{\prime })|^{\lambda _{3}}\omega ^{n})$
and space $O(c_{3}|XSS(u_{i}^{\prime })|^{\lambda _{4}}+c_{1}\omega ^{n})$,
and with additional reasonable assumptions this can be improved to time $%
O(|XSS(u_{i}^{\prime })|^{\lambda _{5}}n^{1+\epsilon })$ and space $%
O(|XSS(u_{i}^{\prime })|^{\lambda _{6}})$ using algorithms to solve the
MSCSP, CDP efficiently, where the element $u_{i}^{\prime }$ is part of the
TTP's public key, $XSS$ is a summit type set and the constant $\omega $
depends on $n$.\medskip

\underline{Proof}\medskip

The following easy computations involved in computing the complexity in the
better case: a randomly chosen generator has probability $p_{\alpha }=\frac{%
l_{\alpha }}{n-2}$ and $p_{\beta }=\frac{r_{\beta }}{n-2}$ of being in $BL$
and $BR$ respectively, then an attacker can selects a random word $b$ from $%
U_{L_{P}}^{\prime }$ using in length of $\theta $ Artin generators then it
has $p_{\alpha ,\beta ,\theta }=\frac{1}{p_{\alpha }^{\theta }}+\frac{1}{%
p_{\beta }^{\theta }}$ probability of being in $BL$ or $BR$. From the
algorithm used to compute $S_{u_{i}^{\prime }}$ (which computes words less
than a certain bound) we do not have to pick $\theta $ too large there exist
some $k$ of short length in Artin generators. Choosing $\theta \leq n$ as
this keeps the algorithm in factorial complexity but this is not a good
choice, from the above discussion the attacker can take $\theta =O(\frac{n}{%
\log (n)})$. Observe the attacker must on average compute 
\begin{equation}
\left\lceil Wp_{\alpha ,\beta ,\theta }^{-1}\right\rceil <\frac{%
(n-2)^{\theta }}{\min (l_{\alpha },r_{\beta })^{\theta }}  \tag{2}
\end{equation}
before expecting $W$ words to be found in $S_{v_{i}^{\prime }}\cup
S_{w_{j}^{\prime }}$. \ The attacker may estimate $p_{\alpha },p_{\beta }$
if the attacker assumes $l_{\alpha }$, $r_{\beta }$, $n$ are not independent
of each other, for example $l_{\alpha }\approx r_{\beta }$ and assumes all
(then $p_{\alpha ,\beta ,1}=1$ for the selection of $\ b$ used in both runs)
or nearly all possible Artin generators are used in $BL,BR$ so $\ p_{\alpha
}\approx p_{\beta }$. Hence (independent of large enough $n$)$,$ the
attacker needs to compute approximately as few as $2^{\theta }$ distinct
words for the parameters suggested in [1] to ensure on average a reduction
to the MSCSP with at least $2$ equations. We would need to select only
approximately $4$ distinct random words of length $3$ from $%
U_{L_{P}}^{\prime }$ before the attacker expects to get one conjugacy
equation or the CSP, the example above use little memory and potentially
little computing power.

In this better case we use an efficient algorithm for the CDP such as the
one given in [6], use a linear search, and use an algorithm for the MSCSP
that works with high probability. \ We assume in all of this second proof,
the length of $b$ is less than $u_{i}$, this means in general $%
O(|SSS(u_{i})|)$ is greater than $O(|SSS(b)|)$.

We assume in this proof there is an algorithm that can compute $%
S_{u_{i}^{\prime }}$ proportional in space and time complexity to $|XSS|$,
this assumption is based on the fact that such an algorithm exists when $%
XSS=SSS$ see [9], and that this algorithm has worst time space and time
exponential complexity. From the description above (from $\left\lceil
Wp_{\alpha ,\beta ,O(\frac{n}{\log (n)})}^{-1}\right\rceil <O(n^{O(\frac{n}{%
\log (n)})})$), the time complexity in this better case is $%
O(|XSS(u_{k}^{\prime })|^{\lambda _{3}}\omega ^{n})$ as shown below. We
assume we use an algorithm for the CDP which has time and space proportional
to $|XSS|$ such as Garside's algorithm [8]. Here $CDP_{t}$ is the average
time taken to solve the CDP over all pairs $(b,a)$. Then by a similar
argument to above the time complexity is 
\begin{eqnarray*}
&&O(S_{t,v_{i}^{\prime }}+S_{t,w_{j}^{\prime }}+CDP_{t}(|(S_{v_{i}^{\prime
}})|+|S_{w_{j}^{\prime }}|)\omega ^{n}+MSCSP_{t}) \\
&\thickapprox &O(E(u_{i}^{\prime })O(n^{O(\frac{n}{\log (n)})})) \\
&=&O(|XSS(u_{k}^{\prime })|^{\lambda _{3}}\omega ^{n})
\end{eqnarray*}
also the space complexity is 
\begin{eqnarray*}
&&O(S_{s,v_{i}^{\prime }}+S_{s,w_{j}^{\prime }}+c_{1}\omega
^{n}+\max_{\forall P\in p_{v},a\in S_{v_{i}^{\prime }}\cup S_{w_{j}^{\prime
}}}\max_{b\in U_{L_{P}}^{\prime }}CDP_{s}(b,a)+MSCSP_{s}) \\
&=&O(c_{2}E(u_{i}^{\prime })+O(n^{O(\frac{n}{\log (n)})})) \\
&\thickapprox &O(c_{3}|XSS(u_{k}^{\prime })|^{\lambda _{4}}+c_{4}\omega ^{n})
\end{eqnarray*}
where the constant $\omega $ depends on the function used for $O(\frac{n}{%
\log (n)})$.\medskip

$\bullet $Better Complexity Bounds\medskip

The above analysis for the better case is may not be optimal, for example if
we make some assumptions then we get a tighter bound on the complexities as
follows. For this case it the average complexities for algorithm for the
CDP,CSP and MSCSP are considered. From the from 2 and if we assume $%
l_{\alpha },r_{\beta }$ are linear in $n$ and $\theta =O(\log (n)),$ we
assume we have an efficient algorithm for the CDP which has average linear
complexity possibly the one given in [5], this assumption is based on the
result that empirically for randomly chosen long random braids which have
simple elements randomly chosen the $|USS|$ is on average likely to be
linear in the word length and independent of the braid index $n$ e.g. see
[5], and use an algorithm for the MSCSP that works with high probability.
Hence the CDP/CSP in this average case can be solved in linear space and
time complexity. The time complexity in this better case can be with high
probability be (recall computing $S_{u_{i}^{\prime }}$ that in proportional
in space an time complexity to $|XSS|$) 
\begin{eqnarray*}
O(c_{1}E(u_{i}^{\prime })\text{\ss }^{O(\log n)}O(n)) &\thickapprox
&O(E(u_{i}^{\prime })n^{1+\epsilon }) \\
&=&O(|XSS(u_{i}^{\prime })|^{\lambda _{5}}n^{1+\epsilon })
\end{eqnarray*}
for some \ss $\in \Re $ which depends on 2. The space complexity in this
better case is 
\begin{eqnarray*}
&&O(c_{1}E(u_{i}^{\prime })+\text{\ss }^{O(\log n)}+S_{s,v_{i}^{\prime
}}+S_{s,w_{j}^{\prime }}) \\
&\thickapprox &O(c_{1}E(u_{i}^{\prime })+c_{2}n^{\epsilon })\thickapprox
O(|XSS(u_{i}^{\prime })|^{\lambda _{6}})
\end{eqnarray*}
using straightforward algebra it can be shown $\epsilon $ can be close to a
constant as $n$ is larger and depends on \ss\ and the constants in $O(\log
n) $, if \ss\ is bounded then $\epsilon $ is bounded.

Note the space can be up to exponential size (so giving a better space bound
here) the only requirement is the set $S_{u_{i}^{\prime }}$ must be of at
least size greater than one as it must contain at least one element \ with
some feasible computable $k$.

The above shows using the AAGL protocol can potentially be as secure than
using CSP based protocols such as the AAG protocol [2] as both can be broken
with attacks of the same or similar complexity depending on the values $%
\lambda _{5}$, $\lambda _{6}$ and $\epsilon $, by similar we mean our
instantiations of our attack can differ by a factor of a polynomial in $n$
from attacks such as on the AAG protocol, for example the time complexities
of attack differ by a factor of $n^{1+\epsilon }$. If the attacker decides
to compute $S_{u_{i}^{\prime }}$ as randomly elements chosen elements of the
centraliser of $u_{i}^{\prime }$ then the success of this attack in this
case depends on the probability of $S_{u_{i}^{\prime }}$ containing elements
of the form $zkz^{-1}$. Informally, or attack consists of computing subset
of centralisers and extracting suitable elements from the centralisers: we
refer to our attack as a two central element attack.

Now consider a variant of the above attack, which is not use an algorithm
for the CDP in step 3iii but instead solves the CSP with the guess for $b$
with every possible element in $S_{u_{i}^{\prime },L_{P}}^{\prime }$ and
recovers $z$ and hence the shared secret using the algorithm in [1], the
attacker may test if $z$ is the correct solution: for example, $z$ is used
in an impersonation attack or if $z^{-1}u_{i}^{\prime }z\sim u_{i}^{\prime }$%
.

Note it may be that $b\in S_{u_{i}^{\prime },L_{P}}^{\prime }$ (which may be
verified using a polynomial time word algorithm in $B_{n}$), in this case $z$
must be in the centraliser of $b$, call the set of all such stored $b,$ $%
b_{z}$ and so $z$ can be found by testing every element of the centraliser\
of a subset of $b_{z}$ for the correct element.

Observe if the attacker assumes his guess of the generators of $BL$ ,$BR$
are correct (or manages know these subgroups in a different way) the
attacker can compute randomly chosen words computable in polynomial time in $%
BL,BR$ and in up to factorial time (in approximately the time taken to solve
the CDP) find a system of conjugacy equation / reduce the security of the
AAGL protocol to the MSCSP so this is another reason why the users should
keep the subgroups $BL$,$BR$ secret. For general $BR$ and $BL$ the algorithm
has to be modified to use the publicly known information about their
structures. The complexity of the algorithm is mainly determined by
computing $S_{v_{i}^{\prime }},S_{w_{j}^{\prime }}$ which may contain
portions of the centralisers, so we can estimate this to be approximately
the same time and space complexity of computing the $SSS$ of an element so
in general it is exponential, also the size of the sets $S_{u_{i}^{\prime
},L_{P}}^{\prime }$ and $U_{L_{P}}^{\prime }$ affect the complexity of the
example algorithm in this connection $O_{1}$ 
\begin{eqnarray*}
O_{1} &=&\sum_{P,\text{ }\forall P\in p_{v}}\sum_{I,\text{ }\forall I\in
i_{v}}|S_{u_{i}^{\prime },L_{P,I}}^{\prime }|+|U_{L_{P,I}}^{\prime
}|+|p_{v}||i_{v}| \\
&=&\sum_{P=P_{0}}^{-2g_{z}}\sum_{I=1}^{f(u_{i}^{\prime })}|S_{u_{i}^{\prime
},L_{P,I}}^{\prime }|+|U_{L_{P,I}}^{\prime }|+f(u_{i}^{\prime })
\end{eqnarray*}
will make the example algorithm can be used as a parameter to measure the
efficiency of algorithm 3, minimising $O_{1}$ will make the algorithm more
efficient. Generally in our probabilistic algorithm we could use an
heuristic optimization algorithm \ instead of a linear search if we do this
then we suggest trying the differential evolution algorithm because it is
known to be fairly fast and reasonably robust [12]. The components of the
vectors used in the differential evolution algorithm depending on $L$ and $%
L_{P}$ the differential evolution means in general the components of the
trial vector will not increase linearly so this means in steps 3ii, 3iv will
not be increased linearly as is done in the probabilistic algorithm.\medskip

\underline{Algorithm-5 To Recover BL and BR.}\medskip

With a little more work we give an attack that recovers the secret subgroups 
$BL$ and $BR$. Any attacker can compute for $i,j$ for sufficiently many $i$
and $j$ using the attack above (to recover $z$) $v_{i},w_{j}$ the attacker
checks for the generator $b_{r}$, $1\leq r\leq n$ if 
\begin{eqnarray}
\text{ }w_{i}b_{r} &=&b_{r}w_{i}\text{ for all }i  \TCItag{3} \\
v_{j}b_{r} &=&b_{r}v_{j}\text{ for all }j\text{.}  \TCItag{4}
\end{eqnarray}
If 3 is true then $b_{r}$ is a generator of $BR$ similarly if 4 is true then 
$b_{r}$ is a generator of $BL$.\medskip

\underline{Algorithm 6-To modify our attack to solve the general MSCSPv.}%
\medskip

1. For the MSCSPv $((v_{1},v_{2},...,v_{u}),(v_{1}^{\prime },v_{2}^{\prime
},...,v_{u}^{\prime }))$ compute a suitable finite approximation $Z$ of the
centralisers of the set of elements $(v_{1}^{\prime },v_{2}^{\prime
},...,v_{u}^{\prime })$.

2. Find elements in $Z$ such that the elements are conjugated by $g$ (we
refer to such elements as the system of conjugacy equations $%
((w_{1},w_{2},...,w_{u}),(w_{1}^{\prime },w_{2}^{\prime },...,w_{u}^{\prime
}))$) such that the sets $(v_{1},v_{2},...,v_{u}),$ $(w_{1},w_{2},...,w_{u})$
are commuting.

3. Solve the MSCSPv using a version of algorithm 2 with the pair of MSCSPv $%
((v_{1},v_{2},...,v_{u}),(v_{1}^{\prime },v_{2}^{\prime },...,v_{u}^{\prime
})),$ $((w_{1},w_{2},...,w_{u}),(w_{1}^{\prime },w_{2}^{\prime
},...,w_{u}^{\prime }))$.\medskip

Here $v_{i}$, $w_{i}$ are chosen from the subgroups $B_{L}$ and $B_{R}$
respectively. Note if the structure of $B_{L}$ and $B_{R}$ is known then
this may be used in our deterministic algorithm. Combinations of centraliser
elements and their inverses of the conjugated generators may be computed to
attempt to construct shorter words $k$. In an example of the above algorithm
it may feasible to compute one or more of the elements$%
\;v_{1},v_{2},...,v_{u}$ possibly using the relation $\lambda (v_{i}^{\prime
})=\lambda (\breve{v}_{i})$ where $\breve{v}_{i}$ is the guess for $v_{i}$,
and hence reducing to the MSCSP there.\medskip

\underline{3.2 Defending Against Attacks}\medskip

The attacks may be avoided if

1) Ensure if possible that elements of the centralisers of $u_{i}^{\prime }$
are hard with the CSP $(k,zkz^{-1})$.

2) Ensure if possible that elements of the centralisers of the form $%
zkz^{-1} $ of $u_{i}^{\prime }$ , that the element $k$ cannot be feasibly
computed.

3) To maximize the value $O_{1}$, with the constraint of making MSCSPv as
difficult as possible for the attacker.

4) The TTP algorithm may be modified with different choices for $BL,BR$ so
that larger generators are used with the constraint of the computing
platform.

5) The security of the AAGL protocol is based on the complexity of algorithm 
$C$ not being efficient and $C$ may be based on the following problem which
is problem 1 given in [10], 
\begin{equation*}
\text{given }g_{1},...,g_{k}\in G\text{ compute }C(g_{1},...,g_{k})
\end{equation*}
here $(g_{1},...,g_{k})=(u_{1}^{\prime },...,u_{\gamma }^{\prime })$ and $%
C(g_{1},...,g_{k})=C(g_{1})\cap C(g_{2})\cap ...C(g_{k})$.\medskip

\underline{4. Potential Length Based Algorithm for MSCSPv}\medskip

We show that modified a known basic length attack for example see [11] can
be used to for the general MSCSPv and then any algorithm can be used to
solve the MSCSP such as a known\ length attack such as [11]. We refer to our
length based MSCSP algorithm the length-MSCSPv algorithm. Suppose we are
given an example of the MSCSPv.

Compute the centralizer of $zw_{i}z^{-1}$ or a portion of this set we call $%
S_{w_{i}^{\prime }}$ then $S_{w_{i}^{\prime }}$ contains all elements of the
form $F=zkz^{-1}$ it follows choices for $k$ includes all elements in $B_{L}$
for a suitable approximation of the centraliser. Hence the generators we
peel from $F$ in our length attack are the generators of the centralisers of 
$w_{i}^{\prime }$. For example in $B_{4}$ one of the generators of the
centraliser of the element $\sigma _{1}^{4}\sigma _{2}\sigma _{3}$ is $%
\sigma _{1}^{2}\sigma _{2}\sigma _{1}\sigma _{3}\sigma _{2}^{-2}\sigma
_{1}^{-3}$ and the above generator is of length 10.\medskip

\underline{Algorithm 7-Length-MSCSPv.}\medskip

Run step 1 and 2 of algorithm 6 for step 3 use the algorithm below instead
of a version of algorithm 2.

Compute 
\begin{equation*}
r_{i}^{\prime
}=zr_{i}z^{-1}zt_{i}z^{-1}zr_{i}^{-1}z^{-1}=zr_{i}t_{i}r_{i}^{-1}z^{-1}
\end{equation*}
where the words $r_{i}$ and $t_{i}\,$\ are a word in the generators $w_{i}$.
Note an element of the form $zr_{i}z^{-1}$ may be used instead for $%
r_{i}^{\prime }$.

1. Select a length function $l$. Construct $r_{i}^{\prime }$ as a word in
the generators $w_{i}^{\prime }$ for some $1\leq i\leq n_{w}$. $\ A$ is set
to the identity element. Set the iteration $n$ to zero. Computes a subset $%
C_{r}$ of the generator set of the intersection of the centralisers
generator sets, $C_{r}\subseteq ($ $C(v_{1}^{\prime })\cap ...\cap
C(v_{a_{c}}^{\prime }))$, $a_{c}=1$ is sufficient, we may also try and
compute the length of the generators of $C_{r}$ with suitably long
generators.

2. Select suitable elements $s_{n}\in C_{r}$. If 
\begin{equation*}
r(r_{1}^{\prime },...,r_{a_{w}}^{\prime },s_{n})\preceq r(r_{1}^{\prime
},...,r_{n_{w}}^{\prime },e)
\end{equation*}
where $\preceq $ is a linear ordering (or an objective function) on a vector
of real numbers [11] and each element of the tuple $r(r_{1}^{\prime
},...,r_{n_{w}}^{\prime },s_{n})$ except the last is given by the
corresponding number $l(s_{n}^{-1}r_{i}^{\prime }s_{n})$.

3. Update the word $A$ as 
\begin{equation*}
A_{n+1}=A_{n}s_{n}\text{.}
\end{equation*}
The algorithm stops at this part when depending on $r(r_{1}^{\prime
},...,r_{n_{w}}^{\prime },s_{n})$ and the stopping criteria (there can be
more than one stopping criteria) then goes to step 6. The algorithm stops
with some probability $\rho $ with $A=z\bar{r}_{i}=\hat{r}_{i}$ and this
braid is stored, where $\bar{r}_{i}=r_{i}C_{i}$ for some $C_{i}\in G$, in
other words $A$ is the product of $z$ and a partial factor of $r_{i}$ we
call $\bar{r}_{i}$.

4. Update the element $r_{i}^{\prime }$ as 
\begin{equation*}
r_{i}^{\prime }=s_{n}^{-1}r_{i}^{\prime }s_{n}
\end{equation*}
5. $n=n+1$. Goto 2.\ 

6. Repeat steps 1 to 5 $a_{w}$ times (obviously with a different choice(s)
elements $r_{i}^{\prime }$ $1\leq i\leq a_{w}$ and maybe a different choice
for the integer $a_{w}$).

7. Steps 1 to 6 are repeated again $a_{v}$ times but with $v_{i}^{\prime }$
in place of $w_{i}^{\prime }$ and $w_{i}^{\prime }$ in place of $%
v_{i}^{\prime }$ using a system of $n_{v}$ equations.

8. We now have stored two sets we refer to as $BV$ and $BW$. 
\begin{eqnarray*}
\{z\bar{w}_{1},z\bar{w}_{2}...,z\bar{w}_{a_{w}}\} &=&\{\hat{w}_{1},...,\hat{w%
}_{a_{w}}\}=BW \\
\{z\bar{v}_{1}\text{,}z\bar{v}_{2}\text{.}..,z\bar{v}_{a_{v}}\} &=&\{\hat{v}%
_{1},...,\hat{v}_{a_{v}}\}=BV\text{ }
\end{eqnarray*}
9. If follows from the MSCSPv example that $\bar{w}_{I}\bar{v}_{J}=\bar{v}%
_{I}\bar{w}_{J}$ for any $I$ or $J$. The attacker picks $I$ and $J$ and
computes 
\begin{equation*}
M_{1}=\hat{v}_{I}^{-1}\hat{w}_{J}=\bar{v}_{I}^{-1}z^{-1}z\bar{w}_{J}=\bar{w}%
_{J}\bar{v}_{I}^{-1}\text{ and }Y=\hat{w}_{J}\hat{v}_{I}^{-1}=z\bar{w}_{J}%
\bar{v}_{I}^{-1}z^{-1}
\end{equation*}
hence the attacker can solves the CSP ($M_{1},Y$) for ($z,z^{-1}$).
Similarly 
\begin{equation*}
M_{2}=\hat{v}_{I}\hat{w}_{J}^{-1}=\bar{v}_{I}z^{-1}z\bar{w}_{J}^{-1}=\bar{w}%
_{J}^{-1}\bar{v}_{I}
\end{equation*}
the CSP an solves the CSP ($M_{2},Y^{-1})$ for $(z,z^{-1}$). Repeating the
above for similar computations for different $I,J$ builds up a system of
conjugacy equations hence this reduces the MSCSPv to the MSCSP.

The algorithm is a modification of known length attack because we use the
generators of the whole conjugated word $zr_{i}z^{-1}$ and not just as usual
the generators of $z$, conjugated element with a partial factor of $r_{i}$
is recovered and intermediate partial factors involving the secret are
recovered and used not as usual the secret element.\medskip

A simple stopping criteria is for some $C$ 
\begin{equation*}
r(t_{1},...,t_{a_{w}},e)<C<r(r_{1}^{\prime },...,r_{a_{w}}^{\prime },e)
\end{equation*}
stop when 
\begin{equation*}
r(r_{i}^{\prime },s_{n})<C
\end{equation*}
and $r(t_{1},...,t_{a_{w}},e)$ is to be estimated by the attacker using the
value of $L$ given in [1].

At step 3 we could solve the equation $z\bar{r}_{i}$ for $z$ which may be
easier than solving the MSCSP instead and so we do not need to run all the
steps, to be precise the algorithm is.\medskip

\underline{Algorithm 8 Length-MSCSPv.}\bigskip

Run step 1 and 2 of algorithm 6 for step 3 use the algorithm below instead
of a version of algorithm 2.

Compute 
\begin{equation*}
r_{i}^{\prime
}=zr_{i}z^{-1}zt_{i}z^{-1}zr_{i}^{-1}z^{-1}=zr_{i}t_{i}r_{i}^{-1}z^{-1}
\end{equation*}
where the words $r_{i}$ and $t_{i}\,$\ are a word in the generators $w_{i}$.
Note an element of the form $zr_{i}z^{-1}$ may be used instead for $%
r_{i}^{\prime }$.

1. \ Select a length function $l$. Construct $r_{i}^{\prime }$ as a word in
the generators $w_{i}^{\prime }$ for some $1\leq i\leq n_{w}$. $\ A$ is set
to the identity element. Set the iteration $n$ to zero. Computes a subset $%
C_{r}$ of the generator set of the intersection of the centralisers
generator sets, $C_{r}\subseteq ($ $C(v_{1}^{\prime })\cap ...\cap
C(v_{a_{c}}^{\prime }))$, $a_{c}=1$ is sufficient, we may also try and
compute the length of the generators of $C_{r}$ with suitably long
generators.

2. Select suitable elements $s_{n}\in C_{r}$. If 
\begin{equation*}
r(r_{1}^{\prime },...,r_{a_{w}}^{\prime },s_{n})\preceq r(r_{1}^{\prime
},...,r_{n_{w}}^{\prime },e)
\end{equation*}
where $\preceq $ is a linear ordering (or an objective function) on a vector
of real numbers [11] and each element of the tuple $r(r_{1}^{\prime
},...,r_{n_{w}}^{\prime },s_{n})$ except the last is given by the
corresponding number $l(s_{n}^{-1}r_{i}^{\prime }s_{n})$.

3. Update the word $A$ as 
\begin{equation*}
A_{n+1}=A_{n}s_{n}\text{.}
\end{equation*}
The algorithm stops at this part when depending on $r(r_{1}^{\prime
},...,r_{n_{w}}^{\prime },s_{n})$ and the stopping criteria (there can be
more than one stopping criteria) then goes to step 6. The algorithm stops
with some probability $\rho _{2}$ with $A=z\bar{r}_{i}=\hat{r}_{i}$ and this
braid is stored, where $\bar{r}_{i}=r_{i}C_{i}$ for some $C_{i}\in G$, in
other words $A$ is the product of $z$ and a partial factor of $r_{i}$ we
call $\bar{r}_{i}$.

4. Update the element $r_{i}^{\prime }$ as 
\begin{equation*}
r_{i}^{\prime }=s_{n}^{-1}r_{i}^{\prime }s_{n}
\end{equation*}
5. $n=n+1$. Goto 2.\ 

6. Repeat steps 1 to 5 $a_{w}$ times (obviously with a different choice(s)
elements $r_{i}^{\prime }$ $1\leq i\leq a_{w}$ and maybe a different choice
for the integer $a_{w}$).

7. Steps 1 to 6 may be repeated again $a_{v}$ times but with $v_{i}^{\prime
} $ in place of $w_{i}^{\prime }$ and $w_{i}^{\prime }$ in place of $%
v_{i}^{\prime }$ using a system of $n_{v}$ equations.

8. We now have stored one set or two sets we refer to as $BV$ and $BW$. 
\begin{eqnarray*}
\{z\bar{w}_{1},z\bar{w}_{2}...,z\bar{w}_{a_{w}}\} &=&\{\hat{w}_{1},...,\hat{w%
}_{a_{w}}\}=BW \\
\{z\bar{v}_{1}\text{,}z\bar{v}_{2}\text{.}..,z\bar{v}_{a_{v}}\} &=&\{\hat{v}%
_{1},...,\hat{v}_{a_{v}}\}=BV\text{ }
\end{eqnarray*}
Using another algorithm we use the elements in $BW$ or $BV$ and solve for $z$
one of the simplest choices at this step is given an element of $BW$ or $BV$
find $\bar{w}_{i}$ or $\bar{v}_{i}$ by brute force and hence compute $z$ by
using a right multiplication.\medskip

\underline{5. Attack Using Conjugacy Extractor Functions}\medskip

\underline{5.1 First Attack using CE Functions}\medskip

In the TTP algorithm above given in [1] step 2 is ``chooses a secret element 
$z\in B_{n}$'' a user could implement this step 2 as $z$ is chosen from a
publicly known subgroup of $B_{n}$ we show that this implementation means a
CE (conjugacy extraction) function [13] can be given. It is not given in [1]
to not pick $z$ from a publicly known subgroup. The attack is as follows.

1. Let $z\in R$ where $R=\{\alpha _{1},...,\alpha _{k}\}$ is a publicly
known subgroup of $B_{n}$. In this step it is required the attacker just
needs to find one element that commutes with $z$ and not with all possible
choices of $u_{i}$ (using a chosen algorithm by the attacker) to show the
AAGL protocol is based on the MSCSP one way to find such elements is as
follows. The attacker picks a subgroup of $R$ given by the generators $%
g_{1},...,g_{k}$. Then the attacker computes all of or a large part of 
\begin{equation*}
S=C(\alpha _{1},...,\alpha _{k})=C(\alpha _{1})\cap ...C(\alpha _{k})\text{.}
\end{equation*}
2.Then 
\begin{equation*}
CE(S_{I},u_{i}^{\prime })=u_{i}^{\prime }S_{I}u_{i}^{\prime
-1}=zu_{i}z^{-1}S_{I}zu_{i}^{-1}z^{-1}=zu_{i}S_{I}u_{i}^{-1}z^{-1},
\end{equation*}
will be true if $S_{I}$ does not commute with $u_{i}$.$\ S_{I}\in S$, $\
1\leq I\leq M$. The the protocol can be based on the MSCSP with 
\begin{equation*}
((S_{1},S_{2},...,S_{M}),(CE(S_{1},u_{i}^{\prime }),CE(S_{2},u_{i}^{\prime
}),...,CE(S_{u},u_{i}^{\prime })))\text{ with solution }(o,o^{-1})\text{, }%
o=zu_{i}.
\end{equation*}
and $z$ can be found by computing $(o^{-1}u_{i}^{\prime })^{-1}=z$.

As a variant of the above algorithm an attacker may try to compute an
element $S_{I}^{^{\prime }}\in C(u_{j}^{^{\prime }})$ then it may be
possible to use $S_{I}^{^{\prime }}$ instead of $S_{I}$ in the attack above
where $u_{j}^{^{\prime }}\neq u_{i}^{^{\prime }}$, so in this variant
knowledge of $z$ being chosen from a subgroup is not required.\medskip

\underline{5.2 Second Attack using CE Functions}\medskip

This attack reveals partial information about the secret $z$.

1. The attacker picks elements $V_{I}$ according to some criteria for
example elements $V_{I}$ may be picked randomly or $V_{I}$ may be composed
of a few Artin generators as these may commute to some degree with $z$.

2. Then for $1\leq I\leq M$ for a sequence of integers $T_{I}$%
\begin{equation*}
CE_{I}(S_{I},u_{T_{I}}^{\prime })=u_{T_{I}}^{\prime }S_{I}u_{T_{I}}^{\prime
-1}=zu_{T_{I}}z^{-1}S_{I}zu_{T_{I}}^{-1}z^{-1}=zu_{T_{I}}\overline{z}%
^{-1}S_{I}\overline{z}u_{T_{I}}^{-1}z^{-1}
\end{equation*}
where $\overline{z}$ is a partial factor of $z$ with probability $\rho _{3}$
for some $I$ this means $z=z_{T_{I}}\overline{z}_{T_{I}}$.

3. We solve for each $I$ the CSP $(S_{I},zu_{T_{I}}\overline{z}^{-1})$ and
hence compute $z_{T_{I}}=((zu_{T_{I}}\overline{z}%
^{-1})^{-1}zu_{T_{I}}z^{-1})^{-1}$

4. We now and find $z$ using the information $(S_{I},zu_{T_{I}}^{-1}%
\overline{z},z_{T_{I}})$ and the other information used in the protocol. One
of the simplest choices to implement this step is trying to find $\overline{z%
}_{T_{I}}$ for each $I$ by brute force.\medskip

A variant of the above attack is after $z_{T_{I}}$ is recovered is to repeat
at the attack (at least once) by iterating with $z_{T_{I}}^{-1}u_{T_{I}}^{%
\prime }z_{T_{I}}$ instead of $u_{T_{I}}^{\prime }$ (and obviously all other
values may be different) so in this way we may be able to find a bigger
factor of $z$. It may be true some probability $\rho _{4}$ that $\overline{z}
$ contains a partial factor of $u_{T_{I}}$ which means the CSP is solved to
give $\overline{z}_{T_{I}}\overline{u}_{T_{I}}$ where $\overline{u}_{T_{I}}$
is some partial factor of $u_{T_{I}}$. Then the simplest choice at this step
to recover $z$ is to find $\overline{u}_{T_{I}}$ by brute force and use $%
\overline{u}_{T_{I}}$ to recover $z$. Note this attack is easily modified to
solve the decompostion problem which means using a product of three elements
instead of $u_{T_{I}}^{\prime }.$

Again another conjugacy extractor (see [13]) (i.e. this will show the AAGL
protocol is based on the MSCSP again). The user may try computations of the
form (following the notation of [1]) $A_{public}\alpha
A_{public}^{-1},A_{public}^{-1}\beta A_{public},B_{public}\gamma
B_{public}^{-1},B_{public}^{-1}\delta B_{public}$ where $A,B,C,D$ are chosen
from $\alpha \in B,\beta \in N_{B},\gamma \in A,\delta \in N_{A}$ . The
information recovered from the MSCSP above may be used in attack such as for
example the following attack. Then once an element of the form may be found $%
z\star (x_{a_{i_{1}}}(t),s_{a_{i_{1}}})\star ...)\star (x_{a_{i_{\mu
}}}(t),s_{a_{i_{\mu }}})\star z^{-1}$ then an element such as $(n_{a},id)$
may be found and so the shared secret can be computed. We will give further
details of this attack. To resist the above attack the public elements
should be chosen so that they do not have an inverse. \medskip

\underline{6. An algorithm for the MSCSP}\medskip

Consider the MSCSP $((x_{1},x_{2},...,x_{u}),(y_{1},y_{2},...,y_{u}))$ with
solution $(g,g^{-1}).$ Suppose $x_{i}\in A,g\in B,$ with $A=\left\langle
a_{1},a_{2},...,a_{M}\right\rangle $ , $B=\left\langle
b_{1},b_{2},...,b_{N}\right\rangle $ \ Compute a large part of all of the
centraliser 
\begin{equation*}
D=C(b_{1},...,b_{N})=C(b_{1})\cap ...C(b_{N})\text{.}
\end{equation*}
The we can compute the $CE$ functions 
\begin{equation*}
CE_{k}(d_{k},y_{i})=y_{i}dy_{i}^{-1}=gx_{i}d_{k}x_{i}^{-1}g^{-1}\text{.}
\end{equation*}
This means we have transformed the MSCSP into another MSCSP. We can use this
transformed MSCSP to attack the protocol in [2], e.g. we may use the
transformed MSCSP as part of another algorithm that solves the MSCSP such as
a length attack, such a length attack is as follows.

1. Select a length function $l$. $\ A$ is set to the identity element. Set
the iteration $n$ to zero. Computes a large part or all of the centraliser $%
D.$

2. While (Criteria=True)

\{

Select elements $s_{n}\in B$.

Compute 
\begin{equation*}
CE_{k}=CE_{k}(d_{k},y_{i})=y_{i}dy_{i}^{-1}=gx_{i}d_{k}x_{i}^{-1}g^{-1}\text{%
.}
\end{equation*}
for some $1\leq k\leq I$ for some $I,d_{k}\in D$. 
\begin{equation*}
r(CE_{1},...,CE_{I},s_{n})\preceq r(CE_{1},...,CE_{I},e)
\end{equation*}
where $\preceq $ is a linear ordering (or an objective function) on a vector
of real numbers [11] and each element of the tuple $%
(CE_{1},...,CE_{I},s_{n}) $ except the last is given by the corresponding
number $l(s_{n}^{-1}CE_{k}s_{n})$.

\}End While.

3. Update the word $A$ as 
\begin{equation*}
A_{n+1}=A_{n}s_{n}\text{.}
\end{equation*}
The algorithm stops at this part when depending on $r(r_{1}^{\prime
},...,r_{n_{w}}^{\prime },s_{n})$ and the stopping criteria (there can be
more than one stopping criteria) then goes to step 6. The algorithm stops
with some probability $\rho _{3}$.

4. Update the element $CE_{k}$ using 
\begin{equation*}
s_{n}^{-1}CE_{k}s_{n}
\end{equation*}
5. $n=n+1$. Goto 2.\ 

6. Output $A$.\medskip

One choice at step 2 is to select $s_{n}$ as choices from all the generators
of $B$, then choice for the Criteria at step 2 is to increase $I$ using a
chosen algorithm until it is decided peeling occurs for one of the $N$
choices (we try all $N$ choices) from $B$ for $s_{n}$: if peeling is still
undecided then the algorithm can pick a generator randomly or stops. We may
include in step 2 the equations $y_{i}$ to peel from in the above.\medskip

\underline{7. Conclusion}\medskip

The above attacks needs to be investigated further, because large parts of
the centraliser for an element can be computed (but in general it is
difficult to compute all elements in the centraliser) and we think the
attacks can be improved. Not considering a brute force algorithm (which is
shown in [1] that the AAGL protocol is secure from a brute force algorithm)
we have given the only deterministic algorithm to break the AAGL protocol.
We have given an algorithms for the MSCSPv is and shown can be reduced to
solving the MSCSP using an algorithm of exponential complexity. Further work
is\smallskip

$\bullet $To implement our deterministic attack or a variant of it for
example, try randomised and/or genetic algorithms (for example these can be
used to increase the probabilities $\rho ,\rho _{i}$), evolutionary
algorithms (e.g. differential evolution) which lead to more probabilistic
solutions (an attacker can try our attack even if it is in worst case of
exponential complexity).

$\bullet $ To minimize $O_{1}$ possibly with additional heuristics in
algorithm 4.

$\bullet $ Try different length attacks apart from the basic length
algorithm (which we have used) in the length-MSCSPv algorithms and to try
different refinements for the above length algorithm these include
randomised and/or genetic algorithms which lead to more probabilistic
solutions. To test/implement the length-MSCSPv algorithms to give
experimental results for its success for different parameters. The
length-MSCSPv algorithms we have given can be used as the basis of other
length-MSCSPv algorithms.

$\bullet $ As described in the attack given in section 5 it is sufficient to
find one element that commutes with $z$ to show the protocol is based on the
MSCSP (and so the AAGL protocol would be no more secure than using another
MSCSP based protocol such as the AAG protocol given in [2]) a natural
question now arises which is.\medskip

Given the example of the MSCSPv used in the AAGL protocol how easy or how
hard is it to find an element $s$ that commutes with $z$ but $s$ does not
commute with all choices of $u_{i}$?\medskip

this question needs further investigation.\smallskip

\underline{References}\medskip

[1] I. Anshel, M. Anshel, D. Goldfeld, S. Lemieux, Key Agreement. The
Algebraic Erasor$^{TM}$ and Light Weight Cryptography in Algebraic Methods
in Cryptography, 418, 2006, AMS

[2] I. Anshel, M. Anshel, D. Goldfeld. An Algebraic Method for Public-key
Cryptography, Mathematical Research Letters, 6, 1999, pp. 287-292.

[3] K.K. Ko, S.J. Lee, J.H. Cheon, J.W. Han, J.S. Kang, C. Park. New
public-key cryptosystem using braid groups. CRYPTO 2000. LNCS, 1880, 2000,
pp. 166-183.

[4] K. H. Ko, Tutorial on Braid Cryptosystems 3, PKC 2001, Korea, February
13-15, 2001. Available at www.ipkc.org/pre\_conf/pkc2001/PKCtp\_ko.ps

[5] V. Gebhardt. A New approach to the conjugacy problem in Garside groups.
J Algebra 292 (1) , 2005, pp. 282-302.

[6] K. Ko, D Choi, M. Cho, J. Lee. New Signature Scheme Using Conjugacy
Problem, available at http://eprint.iacr.org

[7] E. A. Elrifai, H. R. Morton. Algorithms for positive braids, Quart. J.
Math. Oxford., 45, 1994, pp. 479-497.

[8] F. A. Garside. The braid group and other groups, Quart. J. Math. Oxford
78, 1969, pp. 235-254.

[9] N. Franco and J. Gonzalez-Meneses. Computation of centralizers in braid
groups and Garside groups, Rev Mat Iberoamericana 19 (2), 2003, pp. 367-384.

[10] V. Shpilrain and A. Ushakov, A new key exchange protocol based on the
decomposition problem, available at http://eprint.iacr.org/2005/447.pdf
(Accessed 20/05/2006)

[11] D.Garber, S. Kaplan, M. Teicher, B. Tsaban, U. Vishne. Length-Based
Conjugacy Search in the Braid Group, available at
http://arxiv.org/abs/math.GR/0209267

[12] R. Storn, K. Price. Differential Evolution - a Simple and Efficient
Heuristic for Global Optimization over Continuous Spaces, Journal of Global
Optimization, 11, 1997, pp. 341 - 359.

[13] M. M. Chowdhury, On the Security of the Cha-Ko-Lee-Han-Cheon Braid
Group Public-key Cryptosystem, ArXiv preprint, August 2007

[14] P. Dehornoy. Using Shifted Conjugacy in braid based Cryptography,
Contemporary Mathematics.

[15] K. Ko, D Choi, M. Cho, J. Lee. New Signature Scheme Using Conjugacy
Problem, available at http://eprint.iacr.org

[16] J. Lomgrigg and A. Ushakov, Cryptanalysis of Shifted Conjugacy
Authentication Protocol, ArXiv preprint, August 2006\medskip

\underline{Appendix}\medskip

In this appendix we another version algorithm 4 which is presented in the
style of the paper [16]. Then we give an attack on the DSC (Dehornoy Shifted
Conjugacy) protocol in [14]. This appendix follows the style of the LU paper
[16]$.$\bigskip

\underline{ A Probabilistic Algorithm for the Multiple Simultaneous}

\underline{Conjugacy Search Problem Variant}\medskip

We can define the length of the element $x\in B_{n}$ to be the length of its
Garside normal form, and we denote an arbitrary length function by $l(x)$ or 
$l_{i}(x)$.

Recall, we refer the CSP as the MSCSP with $u=1$ so an example of the CSP is
denoted as ($x_{1},y_{1})$. Informally we refer to $x_{1}$ as the ``middle
element'' of $y_{1}$.\medskip

Recall, to break the scheme it is sufficient to find $z$ or a solution that
can be used in place of $z$, then once the common secret conjugate $z$ is
recovered with our attack ,the shared secret key can be computed with the
linear algebraic attack given in [1].\ Recall the security of the protocol
is based on the MSCSPv. The security of the TTP algorithm is based on the
MSCSPv with the elements $(x_{1},x_{2},...,x_{u})=(w_{1},...,w_{\gamma
},v_{1},...,v_{\gamma }),$ $(y_{1},y_{2},...,y_{u})=(w_{1}^{\prime
},...,w_{\gamma }^{\prime },v_{1}^{\prime },...,v_{\gamma }^{\prime })$ and $%
u=2\gamma $. We assume the attacker knows this instance of the MSCSPv in the
Artin representation.

In this appendix we give algorithms which are a probabilistic reduction from
the MSCSPv to the MSCSP this includes another version of algorithm 4.

At the end of the appendix we make suggestion for secure protocols
parameters.\medskip

To summarize our work.\medskip

A. We give an algorithm (another version of algorithm 4) to show the MSCSPv
can be probabilistically reduced to the MSCSP.

B. We give a new algorithm to solve a hard problem (which is a
generalisation of the SCSP) we refer to as the MSSDP.

C. We give an algorithm for the MSSDPv; the MSSDPv generalises the MSCSPv
and the MSSDP simultaneously.\medskip

\textbf{Definition}- $C$ is an algorithm that computes elements in the
centralisers of given elements in factorial space and time complexity in a
worst case.\medskip \newline
\underline{Proposition A}\medskip \newline
For the MSCSPv with $u=1$ if $x_{1}$ can be feasibly computed then $z$ can
be found by solving the CSP($x_{1},y_{1}$).\medskip \newline
\underline{Proof}\medskip \newline
Follows from the definitions of MSCSPv and CSP above.\medskip \newline
\underline{Proposition B }\medskip \newline
Let $u_{i}^{^{\prime }}\sim u_{i}$, for some $1$ $\leq i\leq 2\gamma $ (i.e.
the example of the MSCSPv used in the AAGL protocol) and $z,k\in B_{n}.$
Then for all $i$ we have 
\begin{equation*}
zkz^{-1}\in C_{B_{n}}(u_{i}^{\prime })=C_{B_{n}}(zu_{i}z^{-1})
\end{equation*}
where $C_{B_{n}}(u_{i}^{\prime })$ denotes the centraliser of $u_{i}^{\prime
}$ in $B_{n}$.\medskip \newline
\underline{Proof}\medskip \newline
Obvious.$\square $\medskip \newline
\underline{1 On $n$-Centraliser Attacks}\medskip

Now it follows from the above propositions A, B that the MSCSPv $%
((x_{1},x_{2},...,x_{u}),(y_{1},y_{2},...,y_{u}))$ can be solved in two
steps:\medskip

(S1) Find suitable element(s) $c,$ $c\in C_{B_{n}}(u_{i}^{\prime })$ for at
least two values of $u_{i}^{\prime }$ such that $u_{i}^{\prime }=v_{j}$ and $%
u_{i}^{\prime }=w_{k}^{\prime },$ and $c$ are of the form $zkz^{-1}$ using
algorithm $C$. The computation of the centraliser may be based on the super
summit technique in [7]. We refer to this step as: a $n-$centraliser attack
or a centraliser(s) attack.

(S2) Find using some algorithm: values of $k$, $k=x_{i}$ (for all $i$) in
the MSCSPv then solve the corresponding MSCSP.\medskip

The description of super summit sets is described in [7] so we omit the
description of that part of (S1) here. But (S1) still requires some
elaboration as follows. To be able to work with elements of $%
C_{B_{n}}(u_{i}^{\prime })$ efficiently we need to describe $%
C_{B_{n}}(u_{i}^{\prime })$ in some convenient way, for instance by a set of
generators. Hence (S1) itself consists of two smaller steps: capturing (i.e.
computing suitable approximations of the centraliser(s)) the union of
various centraliser(s) we refer to as $C^{\prime }$, and finding the
required element $c\in C^{\prime }$ in the above union. We formalize the
type of attack in (S1) as follows.\medskip

\textbf{Definition}-An $n-$centraliser attack or a centraliser(s) attack is
an attack where the computation of $n$ centralisers is involved, then a set
of elements from the above $n$ centraliser(s) is found in connection to some
conditions. The elements found above are used as part of the attack. E.g. we
refer to step (S1) as a $(2+d)$-centralisers attack where elements are found
to build an MSCSP, otherwise if $n=1$ we refer to step (S1) as a centraliser
attack or a 1-centraliser attack.\medskip

Our algorithm for the MSCSPv is a $(2+d)$-centraliser attack where $1\leq
d\leq \gamma -1$. A centraliser attack is given in [16] so our idea of
centraliser(s) attack extends the idea of ``a centraliser attack'' in [16].

The only known algorithm [9] for computing a generating set for a
centralizer reduces to the construction of super summit sets, the size of
which is not known to be polynomially bounded, and which is usually hard in
practice. Hence the approach of describing the whole generating set is not
feasible but we will use a variation of this approach. Another approach to
investigate is to find a feasibly computable subgroup as a generating set of
when $k$ is of polynomial length (and hence feasibly computable). By
polynomial above we mean the degree of the polynomial is small enough for
practical computations. We summarize the ideas of this appendix into a
heuristic probabilistic algorithm 1.1 below.

Informally, the algorithm below works because of the following:\medskip

i) If $g_{a}$ is \textit{small} enough then the ``middle elements, `the $%
x_{i}$'s' '' in the MSCSPv can be found by guessing; the ``middle element''
is known in the MSCSP; hence we can reduce MSCSPv to the MSCSP if the above
guess is correct.

ii) $g_{a}$ is suggested to be small in the AAGL protocol because $a$ is
unknown and AAGL is for a lightweight platform.

iii) The structure of the TTP algorithm in the AAGL protocol implies that we
can get $g_{a}=1$ for a suitable choice of the algorithm $C$, and if we
select two values of $u_{i}^{\prime }$ such that $u_{i}^{\prime }=v_{j}$ and 
$w_{k}^{\prime }$ this implies we can inefficiently deterministically reduce
the MSCSPv to the MSCSP. This is because $BL,$ $BR$ do not have any
generators in common and so the middle element must be correct for one of
the two above choices of $u_{i}^{\prime }$.

iv) We can use some type of search, such as a heuristic search, to find the
``middle element'' more efficiently.

v) We can test that we have found the correct ``middle element'' by using
the property of the efficiently computable braid invariant $\lambda $, which
is an invariant in conjugation, i.e. $\lambda (x_{1})=\lambda (y_{1})$ in
the CSP, we can get enough information of the ``middle element'' easily
without solving the triple decompostion problem to get the ``middle
element''.

The text in italics at each step in the probabilistic algorithm is a
suggestion for an example of that step. The algorithm is for a general
example of the MSCSPv but suggestion are made specifically for the AAGL
protocol in our algorithm. Step A implements (S1). Step B implements
(S2).\medskip

\underline{Algorithm 1.1 - Probabilistic algorithm for MSCSPv}\medskip

INPUT: An example of the MSCSPv $(y_{1},y_{2},...,y_{u})$ in $%
((x_{1},x_{2},...,x_{u}),(y_{1},y_{2},...,y_{u}))$ and the value of $u$
called $t$.

OUTPUT: A solution $g^{\prime }$ for the MSCSPv.

COMPUTATION:

\textbf{A}. Set $S=M=C^{\prime }=A=\varnothing $. Using an chosen algorithm $%
C$ compute $C^{\prime }\subset C_{B_{n}}(u_{i}^{\prime })$ for some values
of $i$, i.e. $C^{\prime }=\cup _{\forall i}C_{B_{n}}(u_{i}^{\prime })$. It
follows the $C^{\prime }$ may contain elements of the form $F=zkz^{-1}$.
Hence for choices of $k$ includes all elements in $BR$ or $BL$ depending if $%
u_{i}^{\prime }=v_{i}^{\prime }$ or $u_{i}^{\prime }=w_{i}^{\prime }$.

\textit{There are two choices we suggest for this step:}

First choice\textit{: compute the centraliser as a generating set using the
algorithm in }[9]\textit{. Then select random products of the generators to
give a word }$r$\textit{. An option for this choice is using a suitable
length function }$l_{1}$\textit{\ we would expect if r is conjugated by z } 
\begin{equation*}
l_{1}(ru_{i}^{\prime })<l_{1}(r)+l_{1}(u_{i}^{\prime })\text{ and }%
l_{1}(u_{i}^{\prime }r)<l_{1}(r)+l_{1}(u_{i}^{\prime })
\end{equation*}
The above idea is based on the Hamming distance between words: i.e. if $r$
and $u_{i}$ are both conjugated by the same element then we would expect
both the above inequalities to be true.

Second choice\textit{: Use the subalgorithm for this step described below.}

\textbf{B} Repeat steps Bi, Bii, Biii until a solution is found.

We construct the pair $(a,zaz^{-1})=(a,b)$, find $a$ as follows, and $%
zaz^{-1}\in C^{\prime }$ as follows.

Bi. Select $b\in C^{\prime }\backslash M$ and add $b$ to $M$ if $b$ is
selected\textit{. }If all possible values for $b$ have been used (i.e. $%
M=C^{\prime }$) goto step A. Using a chosen algorithm we find $b$ in the
pair $(a,b)$ as follows. \textit{A choice at this step is that because
anyone knows the length }$g_{z}$\textit{\ in Artin generators of }$z$\textit{%
\ [1] and we assume the length in Artin generators of }$a$\textit{\ is less
a feasible bound }$g_{a}$\textit{, hence we do a type of search (e.g. linear
search or random search) using an algorithm that uses }$l_{2}(b),g_{z},g_{a}$%
\textit{\ e.g. an algorithm that uses the heuristic } 
\begin{equation*}
\mathit{l}_{2}\mathit{(b)\leq 2g}_{z}\mathit{+g}_{a}\text{.}
\end{equation*}
Bii. Construct using a chosen algorithm, the set $A$ which is the set of
possible values for $a$. Then using a chosen algorithm we find $a\in A$. If
the relation $b\sim a$ is true then add the pair $(a,b)$ to $S$, because
this means $a$ is conjugate to $b$. $M$ represents worked out elements for $%
b $. $S$ represents worked out elements that are used in an MSCSP. \textit{A
choice at this step is that we can try out all possible values of }$a$%
\textit{\ up to length }$g_{a}$\textit{\ (using a chosen length function)
this includes easy to guess choices which exist (follows from the TTP
algorithm) for }$a$\textit{\ which are single Artin generators. We then use
the practical algorithm for the CDP in [6] to test the relation }$b\sim a$%
\textit{\ } 
\begin{equation*}
\mathit{\lambda (b)\sim \lambda (a)}
\end{equation*}
\textit{where }$\lambda $\textit{\ is a braid invariant. }

iii. Repeat steps i and ii until the desired value of $u=t$ is reached in
the MSCSP. (If the desired value of $t$ not reached then at $i$ goes step
A). Otherwise goto step C.

\textbf{C}. Solve the MSCSP for all the pairs $(a,b)$ using an algorithm
that works with high probability: if solution has been found terminate
algorithm, if the solution of the MSCSP has not been found goto step
A.\medskip

\underline{Algorithm 1.2-Second choice for Subalgorithm in Step A of
algorithm 1.1}\medskip

Here we can use a optimization method (e.g. simulated annealing) by
considering the $(y_{1},y_{2},...,y_{u})$ as input to the optimization
method and minimising the length (using a chosen length function) of $a$ in
the CSP pair $(a,b).$ For example the simplest choice here is the following
simple optimization algorithm. The idea of this subalgorithm is words formed
from words in the generators $(y_{1},y_{2},...,y_{u})$ are in $C^{\prime }$,
and such word(s) may have a small length in $a$. The subalgorithm can be
used at least twice depending on if $u_{i}^{\prime }=v_{i}^{\prime }$ or $%
u_{i}^{\prime }=w_{i}^{\prime }$.\medskip

1. Initialisation step. Choose a suitable length function $l$. Set $%
fitness=\min_{All\text{ k, 1}\leq k\leq u}l(y_{k})$. Set $Solution=l(y_{k})$
with $y_{k}$ such that it has minimal fitness (i.e. it is a solution).

2. Select random subsets of the MSCSPv $(y_{1},y_{2},...,y_{u})$ tuple, the
simplest choice at this step is to select random $y_{i},y_{j}$ for two
random $i$ and $j$.

3. The objective function should grow smaller as $a$ is smaller. Then the
simplest choice is using 
\begin{equation*}
Objective\text{ }function=l(y_{i}^{\pm 1}y_{j}^{\pm 1})\text{.}
\end{equation*}
Then if $l(y_{i}^{\pm 1}y_{j}^{\pm 1})<fitness$ we add the element $%
y_{i}^{\pm 1}y_{j}^{\pm 1}$ to $C^{\prime }$ if $y_{i}^{\pm 1}y_{j}^{\pm
1}\notin C^{\prime }$. Where $l$ is a length function.

4. Repeat steps 2 and 3 until the desired until the desired size of $%
|C^{\prime }|$ is reached. If the desired size is not reached then proceed
to the next step in the main algorithm.\medskip \newline
\underline{Proposition 1.1}\medskip \newline
Let $\lambda _{1},\lambda _{2},\omega ,c_{1},c_{2}\in \Re $. Solving the
MSCSPv can be done with the probabilistic algorithm 1.1, in approximately,
time complexity $O(|XSS(u_{i}^{\prime })|^{\lambda _{2}}\omega ^{n})$ and
space complexity $O(c_{1}\omega ^{n}+c_{2}|XSS(u_{i}^{\prime })|^{\lambda
_{1}})$.\medskip \newline
\underline{Proposition 1.2}\medskip \newline
Let $\lambda _{1},\lambda _{2},\omega ,c_{1},c_{2}\in \Re $. This
proposition improves the complexity bound of proposition 1.1: with
additional reasonable assumptions it can be improved to time complexity $%
O(|XSS(u_{i}^{\prime })|^{\lambda _{1}}n^{1+\epsilon })$ and space
complexity $O(|XSS(u_{i}^{\prime })|^{\lambda _{2}})$ using algorithms to
solve the MSCSP, CDP efficiently, where the element $u_{i}^{\prime }$ is
part of the TTP's public key, $XSS$ is a summit type set and the constant $%
\omega $ depends on $n$.\medskip \newline
\underline{Proof for Proposition 1.1}\medskip

We assume that algorithm 1.1 is successful this means it gives the correct
output and in particular at step Biii the value of $t=u$ is reached and that
a linear search is used at Bi. We do the complexity analyses by following
the algorithm 1 through a successful execution.

\textbf{Step A}. The complexity at step A is determined by the algorithm $%
C^{\prime }$ which computes elements in the centraliser. It is assumed that $%
C^{\prime }$ has factorial space and time complexity: this assumption is
based on the fact that such algorithms exists e.g. when $XSS=SSS,$ or $%
XSS=SS $ see [9], [8] ; precisely we mean $C^{\prime }$ has space and time
complexity proportional to $|XSS(u_{i}^{\prime })|$. Recall, generically an
element of $C^{\prime }$ contains elements of the form $F=zaz^{-1};$ this is
an important thing to observe.

\textbf{Step Bi}. Since we are trying to construct an MSCSP in $t$ equations
it follows we have to do step Bi at least $t$ times (i.e. at step Biii we
repeat $t$ times). In the worst case we would have to try out all (i.e.
using the linear search) of the elements of $C^{\prime }$ (stored in the
above step) which is of size $O(|XSS(u_{i}^{\prime })|^{\lambda _{1}})$.

\textbf{Step Bii}. Recall we are trying to find $a$ in $F$. We analyse the
simplest method which is simply to randomly guess $a$ in its Artin
generators. This gives the following straightforward computations.

$\bullet $ From the TTP algorithm it follows a randomly chosen generator has
probability $p_{\alpha }=\frac{l_{\alpha }}{n-2}$ and $p_{\beta }=\frac{%
r_{\beta }}{n-2}$ of being in $BL$ and $BR$ respectively. Hence an attacker
can selects a random word $a$ from $A$ using in length of $g_{a}$ Artin
generators then it has $p_{\alpha ,\beta ,g_{a}}=\frac{1}{p_{\alpha }^{g_{a}}%
}+\frac{1}{p_{\beta }^{g_{a}}}$ probability of being in $BL$ or $BR$; the
above is true because we selected above values of $u_{i}^{\prime }$ such
that $u_{i}^{\prime }=v_{j}$ and $w_{k}^{\prime }$. From step Bi we compute
a subset of $C^{\prime }$ (with words less than a certain bound) we do not
have to pick $g_{a}$ too large because as noted above there exist some $a$
of short length in Artin generators.

$\bullet $ Observe the attacker must on average compute 
\begin{equation}
\left\lceil Wp_{\alpha ,\beta ,g_{a}}^{-1}\right\rceil <\frac{(n-2)^{g_{a}}}{%
\min (l_{\alpha },r_{\beta })^{g_{a}}}  \tag{a}
\end{equation}
before expecting $W$ words to be found in for $a$ in $(a,zaz^{-1})\in
(A,C_{B_{n}}(u_{i}^{\prime }))$.

Choosing $g_{a}\leq n$ keeps the algorithm in factorial complexity but this
is not a good choice, from the above discussion the attacker can take $%
g_{a}=O(\frac{n}{\log (n)})$; in particular from $\left\lceil Wp_{\alpha
,\beta ,O(\frac{n}{\log (n)})}^{-1}\right\rceil <O(n^{O(\frac{n}{\log (n)}%
)}))=O(e^{n})=\omega ^{n}$ (for some $\omega \in \Re ),$ means this part is
exponential.

$\bullet $\ It follows the attacker needs to estimate $p_{\alpha },p_{\beta
} $. The attacker may estimate $p_{\alpha },p_{\beta }$ if the attacker
assumes $l_{\alpha }$, $r_{\beta }$, $n$ are not independent of each other,
for example $l_{\alpha }\approx r_{\beta }$ and assumes for example, all
nearly all possible Artin generators are used in $BL,BR:$ so $\ p_{\alpha
}\approx p_{\beta }$. Note if all possible Artin generators are used then $%
p_{\alpha ,\beta ,1}=1,$ for the selection of$\ a$ (the single Artin
generator). Hence (independent of large enough $n$ and when $l_{\alpha
}\approx r_{\beta }$)$,$ the attacker needs to compute approximately as few
as $2^{g_{a}}$ distinct words for the parameters suggested in [1] to ensure
on average a reduction to the MSCSP with at least $2$ equations. So we would
need to select only approximately $4$ distinct random words of length $3$
from the set of possibilities of $k$, before the attacker expects to get one
conjugacy equation or the CSP, the example above use little memory and
potentially little computing power.

At the next part in step Bii we use an efficient algorithm for the CDP such
as the one given in [6], use a linear search, and use an algorithm for the
MSCSP that works with high probability.\ It is expected from Bi the length
of $b$ is less than $u_{i}$, this means in general $O(|SSS(u_{i})|)$ is
greater than $O(|SSS(b)|)$.

\textbf{Step Biii}. Steps Bi to Bii are repeated $t$ times hence the
complexity in the steps is has a factor of $t$. \medskip

We can now evaluate the total complexity of the algorithm.\textbf{\medskip }

\textbf{Notation}-The notation $A_{s}$ and $A_{t}$ means the space
complexity and time complexity respectively for an arbitrary algorithm
labelled $A$.\medskip

\underline{The time complexity is in the worst case is} 
\begin{eqnarray*}
&&O(C_{B_{n}}(u_{i}^{\prime })_{t}+CDP_{t}\cdot |C_{B_{n}}(u_{i}^{\prime
})|\cdot \omega ^{n}+MSCSP_{t}) \\
&&\text{hence an upper bound is }O(|XSS(u_{k}^{\prime })|^{\lambda
_{1}}\omega ^{n}).
\end{eqnarray*}
The explanation is as follows. The term $C_{B_{n}}(u_{i}^{\prime })_{t}$
used in step A is between $2$ and $2+d$ times or a constant number of times,
and this implies the corresponding constant in the complexity term can be
ignored. The term $MSCSP_{t}$ is the complexity at the last step and has
order less than $|XSS(u_{k}^{\prime })|^{\lambda _{1}}$. The term $%
CDP_{t}\cdot |C_{B_{n}}(u_{i}^{\prime })|\cdot \omega ^{n}$ is the
complexity at step Bii: at step Biii means step Bii is repeated $\omega ^{n}$
times for each possible value of $a$ in conjunction with each possible value
for $b$; there are $C_{B_{n}}(u_{i}^{\prime })_{s}$ values for $b$. Here the
constant $CDP_{t}$ is the average time taken to solve the CDP over all pairs 
$(a,b)$. In the worst case space complexity at this sub step is $%
O(t|XSS(u_{i}^{\prime })|^{\lambda _{1}})$. Clearly the term $CDP_{t}\cdot
|C_{B_{n}}(u_{i}^{\prime })|\cdot \omega ^{n}$ dominates the complexity.
Because the $CDP$ is done using an efficient probabilistic algorithm such as
[6] and $t\leq \omega ^{n}$ the time complexity follows.

\underline{The worst space complexity is} 
\begin{eqnarray*}
&&O(|C_{B_{n}}(u_{i}^{\prime })|+|A|+\max_{a\in A}\max_{b\in C^{\prime
}}CDP_{s}(a,b)+MSCSP_{s}) \\
&&\text{ hence an upper bound is }O(c_{1}|XSS(u_{k}^{\prime })|^{\lambda
_{2}}+c_{2}\omega ^{n}).
\end{eqnarray*}
The explanation is as follows. The terms $|C_{B_{n}}(u_{i}^{\prime
})|,MSCSP_{s}$ both have complexity equal or less than to $%
O(c_{1}|XSS(u_{k}^{\prime })|^{\lambda _{2}})$ (so we have combined both
complexities into one term). $|A|$ is of size $c_{2}\omega ^{n}$. $%
\max_{a\in A}\max_{b\in C^{\prime }}CDP_{s}(a,b)$ is negligible as the
practical algorithm [6] stores only two elements. $\square .$

Now we are in a position to prove proposition 1.2.\medskip

\underline{Proof for Proposition 1.2}\medskip

The above analysis may not be optimal, for example if we make some
reasonable assumptions then we get a better bound on the complexities as
follows. For this case it the average complexities (instead of worst case)
for algorithm for the CDP,CSP and MSCSP are considered.

The reasonable assumptions we are as follows.

$\bullet $ From [1] we assume $l_{\alpha },r_{\beta }$ are linear in $n$ and
the attacker selects $g_{a}=O(\log (n))$.

$\bullet $ We assume we have an efficient algorithm for the CDP which has
average linear complexity possibly the one given in [5], this assumption is
based on the result that empirically for randomly chosen long random braids
which have simple elements randomly chosen, the $|USS|$ is on average likely
to be linear in the supremum and independent of the braid index $n$ e.g. see
[5]. Hence the CDP/CSP in this average case can be solved in linear space
and time complexity;

$\bullet $ use an algorithm for the MSCSP that works with high probability
such as the one in [5].\medskip

\underline{It follows (using the assumptions) the dominant term in the time
complexity for this term} is the dominant term for the time complexity in
the related proposition 1.1 with the term \ss $^{O(\log n)}$ replacing $%
\omega ^{n}$. The time complexity in this better case can be with high
probability be (recall computing $C^{\prime }$ that in proportional in space
an time complexity to $|XSS|$) 
\begin{equation*}
O(c_{1}|XSS(u_{i}^{\prime })|^{\lambda _{1}}\text{\ss }^{O(\log
n)}O(n))=O(|XSS(u_{i}^{\prime })|^{\lambda _{1}}n^{1+\epsilon })
\end{equation*}
for some \ss $\in \Re $ which depends on equation a. The factor $O(n)$ is
for the complexity for the $CDP$ algorithm.

\underline{It follows (using the assumptions) the dominant term in the space
complexity for this term} is related to the space complexity in the related
case above proposition 1.2. The space complexity is 
\begin{eqnarray*}
&&O(c_{1}|XSS(u_{i}^{\prime })|^{\lambda _{2}}+\text{\ss }^{O(\log
n)}+|C_{B_{n}}(u_{i}^{\prime })|) \\
&\thickapprox &O(c_{1}|XSS(u_{i}^{\prime })|^{\lambda _{2}}+c_{2}n^{\epsilon
})\thickapprox O(|XSS(u_{i}^{\prime })|^{\lambda _{2}})\text{.}
\end{eqnarray*}
Using straightforward algebra it can be shown $\epsilon $ can be close to a
constant as $n$ becomes larger and depends on \ss\ and the constants in $%
O(\log n)$, if \ss\ is bounded then $\epsilon $ is bounded.

Note the space can be up to exponential size (so giving a better space bound
here), the only requirement for the algorithm to successfully terminate, is
the set $C^{\prime }$ must be non-empty than one as it must contain at least
one element with some feasible computable $k.$

The above shows using the AAGL protocol can potentially be as secure as
using CSP based protocols such as the AAG protocol [2] as both can be broken
with attacks of the same or similar complexity depending on the values $%
\lambda _{1}$, $\lambda _{2}$ and $\epsilon $, by similar we mean our
instantiations of our attack can differ by a factor that is polynomial of
the bitlength attack input (i.e. similar) from attacks such as on the AAG
protocol, for example the time complexities of attack differ by a factor of $%
n^{1+\epsilon }$ compared to the $SSS$ attack.

We can try a variant of the above algorithm, which is not use an algorithm
for the CDP in step Bii but instead solves the CSP with the guess for $b$
with every possible element in $C^{\prime }$ and recovers $z$, and hence the
shared secret using the linear algebraic attack given in [1], the attacker
may test if $z$ is the correct solution: for example by computing if $%
z^{-1}u_{i}^{\prime }z\sim u_{i}^{\prime }$.

Another variant we can try is: because it may be that $b\in C^{\prime }$
(which may be verified using a polynomial time word algorithm in $B_{n}$),
in this case $z$ must be in the centraliser of $b$, call the set of all such
stored $b$ as $b_{z},$ and so $z$ can be found by testing every element of
the centraliser\ of a subset of $b_{z}$ for the correct element.$\square $

Observe if the attacker assumes his guess of the generators of $BL$, $BR$
are correct (or manages know these subgroups in a different way) the
attacker can compute randomly chosen words computable in polynomial time in $%
BL,BR,$ and in up to factorial time (in approximately the time taken to
solve the CDP) find a system of conjugacy equation / reduce the security of
the AAGL protocol to the MSCSP, so this is another reason why the users
should keep the subgroups $BL$,$BR$ secret. For the modification to the AAGL
which is using general $BR$ and $BL$ the algorithm, then our attack has to
be modified to use the publicly known information about the structures of $%
BL $ and $BR$.

We see to increase the probability of for our attack on the AAGL protocol
above to succeed we have to compute a ``something like a geodesic of $%
u_{i}^{\prime }$''. A geodesic of a braid is a braid word of minimum length
in the Artin generators representing a given braid. It is known that
computing the geodesic of a braid is an NP-complete problem. However the
version of the geodesic problem the AAGL protocol is based upon is to find a
word equivalent to $zaz^{-1}$ such that it is short enough for $a$ to be
feasible computed. WLOG assume $za\in B_{n}^{+}$, the above problem is
(easily) equivalent to replacing $zaz^{-1}$ by $zaz^{\ast }$ where $z^{\ast
}=z^{-1}\Delta ^{\sup (z)}$, so it is sufficient to find a geodesic like
element of $zaz^{\ast }$ in $B_{n}^{+}$. Even though computing a geodesic is
NP-complete there are two reasons why the above problem may still be easy:
the first is the problem is a version of the geodesic problem and not the
exact geodesic problem; the second is there are many NP-complete problems
have polynomial time average case solutions, e.g. observe it is easy compute
a geodesic of a permutations braid hints that there is such a solution of
the above problem.\medskip

\underline{2 A Centralisers Attack on the Multiple Simultaneous Shifted
Decomposition Problem -MSSDP}\medskip

Recall the shift operator in $B_{\infty }$ for the word $w=\sigma
_{i_{1}}^{\epsilon _{1}}...\sigma _{i_{k}}^{\epsilon _{k}}$ as the word 
\begin{equation*}
d(w)=\sigma _{i_{1+1}}^{\epsilon _{1}}...\sigma _{i_{k}+1}^{\epsilon _{k}}
\end{equation*}
This operator induces a monomorphism on the infinite braid group. Recall the
braid $a\ast b$ is 
\begin{equation*}
a\ast b=a\cdot d(b)\cdot \sigma _{1}\cdot d(a^{-1}),
\end{equation*}
and the operator $\ast $ is the shifted conjugacy operator. Recall the SCSP
(shifted conjugacy search problem) is defined as the following hard problem.
For braids $x,y,c\in B_{\infty }$ find a braid $x\in B_{\infty }$ such that $%
y=x\ast c:$ where $c,y$ are publicly known and $x$ is secret.

We now generalise the SCSP in a straight forward way to a hard decomposition
type problem called the SDP.

\textbf{Definition}. The SDP (shifted decomposition problem), for braids $%
w,x,y,c\in B_{\infty }$ find braids $w,x\in B_{\infty }$ such that $y=w\cdot
d(c)\cdot \sigma _{1}\cdot d(x)$ where $c,y$ are publicly known and $w,x$
are secret. The SDP is a generalisation because with $x=w^{-1}$ we recover
the SCSP.

\textbf{Notation}. We use the notation $w\ast c\ast x=y$ for the SDP.

\textbf{Definition}. The MSSDP (multiple simultaneous SDP), is a set of SDP
equations, as follows. Let $n\geq 1$ be a fixed integer. For braids $%
w,x,y_{i},c_{i}\in B_{\infty },$ $1\leq i\leq n$ find braids $w,x\in
B_{\infty }$ such that $y_{i}=wd(c_{i})\sigma _{1}d(x)$ where $c_{i},y$ are
publicly known and $w,x$ are secret.\medskip

There are no efficient solutions for solving the MSSDP, one reason for is
this would mean the SCSP would be easy. We propose a solution for MSSDP.

Consider $n=4$ in the MSSDP and $c_{1}\neq c_{2}$, $c_{3}\neq c_{4}$, this
is the system of equations 
\begin{eqnarray}
y_{1} &=&wd(c_{1})\sigma _{1}d(x),\text{ }y_{2}=wd(c_{2})\sigma _{1}d(x) 
\TCItag{b} \\
y_{3} &=&wd(c_{3})\sigma _{1}d(x)\text{ and }y_{4}=wd(c_{4})\sigma _{1}d(x)%
\text{.}  \notag
\end{eqnarray}
We now use the idea CE (conjugacy extractor) [13] used to transform the
MSSDP into a shifted MSCSP type problem very efficiently: the transformation
is achieved using CE functions, the concept of CE functions were first
introduced in our paper [13]. A CE function is defined as follows, \textbf{%
definition} A CE function uses input from public information in a hard
problem and transforms the hard problem into a an example of the CSP. First
we give the mathematical background then we give our centralisers
attack.\medskip

\underline{On an algorithm For the MSSDP}\medskip

Define the CE function for one SDP in b as 
\begin{equation}
CE(y_{1},y_{2})=y_{1}^{-1}y_{2}=d^{-1}(x)\sigma
_{1}^{-1}d^{-1}(c_{1})d(c_{2})\sigma _{1}d(x)  \tag{c}
\end{equation}
which is equivalent to solving the CSP with $(\sigma
_{1}^{-1}d^{-1}(c_{1})w^{-1}wd(c_{2})\sigma _{1},CE(y_{1},y_{2}))$ with
solution $d(x)$.

For $n\in N$ define the braids 
\begin{equation*}
\delta _{n}=\sigma _{n-1}...\sigma _{1}.
\end{equation*}
Then for $i=1,...,n-1$%
\begin{equation}
\delta _{n+1}^{-1}\sigma _{i}\delta _{n+1}=_{B_{n+1}}\sigma _{i+1}=d(\sigma
_{i})\text{.}  \tag{d}
\end{equation}
\underline{Proposition 2.1}\medskip

Let $x,CE(y_{1},y_{2}),\sigma _{1}^{-1}d^{-1}(c_{1})d(c_{2})\sigma _{1}\in
B_{n}$. Then $d(x)$ satisfies equation c for the CSP $(\sigma
_{1}^{-1}d^{-1}(c_{1})d(c_{2})\sigma _{1},CE(y_{1},y_{2}))$ if and only if
it satisfies the CSP $(\delta _{n+1}\sigma
_{1}^{-1}d^{-1}(c_{1})d(c_{2})\sigma _{1}\delta _{n+1}^{-1},\delta
_{n+1}CE(y_{1},y_{2})\delta _{n+1}^{-1})$ i.e. 
\begin{equation}
\delta _{n+1}CE(y_{1},y_{2})\delta _{n+1}^{-1}=x\delta _{n+1}\sigma
_{1}^{-1}d^{-1}(c_{1})d(c_{2})\sigma _{1}\delta _{n+1}^{-1}x^{-1}\text{.} 
\tag{e}
\end{equation}

\underline{Proof}\medskip

Follows from d.\medskip

\underline{Proposition 2.2}\medskip

let $x,CE(y_{1},y_{2}),\sigma _{1}^{-1}d^{-1}(c_{1})d(c_{2})\sigma _{1}\in
B_{n}$ be braids satisfying equation c and let $x_{1}^{\prime }\in B_{n+1}$.
Then 
\begin{eqnarray*}
\delta _{n+1}CE(y_{1},y_{2})\delta _{n+1}^{-1} &=&x_{1}^{\prime }\delta
_{n+1}\sigma _{1}^{-1}d^{-1}(c_{1})d(c_{2})\sigma _{1}\delta
_{n+1}^{-1}x_{1}^{\prime -1}\Leftrightarrow \\
x_{1}^{\prime -1}x &\in &C_{B_{n+1}}(\delta _{n+1}\sigma
_{1}^{-1}d^{-1}(c_{1})d(c_{2})\sigma _{1}\delta _{n+1}^{-1})
\end{eqnarray*}
where $C_{B_{n+1}}$ is a centraliser in $B_{n+1}$.\medskip

\underline{Proof}\medskip

The proof for the similar proposition in [16] is ``obvious'' and so is the
proof for this proposition.

Now consider the equation 
\begin{equation}
CE(y_{3},y_{4})=y_{3}^{-1}y_{4}=d^{-1}(x)\sigma
_{1}^{-1}d^{-1}(c_{3})d(c_{4})\sigma _{1}d(x)  \tag{f}
\end{equation}
We can now easily derive two very similar propositions to 2.1, 2.2 where we
use $y_{3},y_{4}$ in place of $y_{1},y_{2}$ respectively. To be precise and
to be complete the propositions are 2.3 and 2.4.\medskip

\underline{Proposition 2.3}\medskip

Let $x,CE(y_{3},y_{4}),\sigma _{1}^{-1}d^{-1}(c_{3})d(c_{4})\sigma _{1}\in
B_{n}$. Then $d(x)$ satisfies the equation f for the CSP $(\sigma
_{1}^{-1}d^{-1}(c_{3})d(c_{3})\sigma _{1},CE(y_{3},y_{4}))$ if and only if
it satisfies the CSP $(\delta _{n+1}\sigma
_{1}^{-1}d^{-1}(c_{3})d(c_{4})\sigma _{1}\delta _{n+1}^{-1},\delta
_{n+1}CE(y_{3},y_{4})\delta _{n+1}^{-1})$ i.e. 
\begin{equation}
\delta _{n+1}CE(y_{3},y_{4})\delta _{n+1}^{-1}=x^{-1}\delta _{n+1}\sigma
_{1}^{-1}d^{-1}(c_{3})d(c_{4})\sigma _{1}\delta _{n+1}^{-1}x\text{.}  \tag{g}
\end{equation}
\medskip

\underline{Proposition 2.4}\medskip

Let $x,CE(y_{3},y_{4}),\sigma _{1}^{-1}d^{-1}(c_{3})d(c_{4})\sigma _{1}\in
B_{n}$. be braids satisfying f and let $x_{2}^{\prime }\in B_{n+1}$. Then 
\begin{eqnarray*}
\delta _{n+1}CE(y_{3},y_{4})\delta _{n+1}^{-1} &=&x_{2}^{\prime }\delta
_{n+1}\sigma _{1}^{-1}d^{-1}(c_{3})d(c_{4})\sigma _{1}\delta
_{n+1}^{-1}x_{2}^{\prime -1}\Leftrightarrow \\
x_{2}^{\prime -1}x &\in &C_{B_{n+1}}(\delta _{n+1}\sigma
_{1}^{-1}d^{-1}(c_{3})d(c_{4})\sigma _{1}\delta _{n+1}^{-1})
\end{eqnarray*}
We now describe a 2-centralisers attack on the MSSDP that recovers $x$. Once 
$x$ is recovered we attempt to find $w$ by computing $y_{i}(d(c_{i})\sigma
_{1}d(x))^{-1}=^{?}w.$

Now it follows from the above four propositions the MSSDP can be solved
using the following steps.

(S1). Find the solution $x_{1}^{\prime },x_{2}^{\prime }\in B_{n+1}$ 
\begin{eqnarray}
\delta _{n+1}CE(y_{1},y_{2})\delta _{n+1}^{-1} &=&x_{1}^{\prime }\delta
_{n+1}\sigma _{1}^{-1}d^{-1}(c_{1})d(c_{2})\sigma _{1}\delta
_{n+1}^{-1}x_{1}^{\prime -1}  \TCItag{h} \\
\delta _{n+1}CE(y_{3},y_{4})\delta _{n+1}^{-1} &=&x_{2}^{\prime }\delta
_{n+1}\sigma _{1}^{-1}d^{-1}(c_{3})d(c_{4})\sigma _{1}\delta
_{n+1}^{-1}x_{2}^{\prime -1}  \TCItag{i}
\end{eqnarray}
this can be done using and $XSS$ based algorithm e.g. using the $USS$
technique of [5].

(S2). ''Correct'' the elements $x_{1}^{\prime },x_{2}^{\prime }\in B_{n+1}$
to obtain a solution to get $s\in B_{n}$ for the MSCSP in c and f i.e. find
elements $C_{1},C_{2}$ such that 
\begin{equation*}
C_{1},C_{2}\in C_{B_{n+1}}(\delta _{n+1}\sigma
_{1}^{-1}d^{-1}(c_{1})d(c_{2})\sigma _{1}\delta _{n+1}^{-1})\cup
C_{B_{n+1}}(\delta _{n+1}\sigma _{1}^{-1}d^{-1}(c_{3})d(c_{4})\sigma
_{1}\delta _{n+1}^{-1})
\end{equation*}
to obtain a solution 
\begin{equation}
t=x_{1}^{\prime }C_{1}=x_{2}^{\prime }C_{2}\in B_{n}\text{.}  \tag{j}
\end{equation}
In j we are using the fact that we are solving an MSCSP c, f and attempting
to recover the same value $x$ (i.e. $x=t$ here) in that MSCSP. This step is
a 2-centralisers attack.

We now derive a feasibly computable subgroup of $C_{B_{n+1}}(\delta
_{n+1}\sigma _{1}^{-1}d^{-1}(c_{1})d(c_{2})\sigma _{1}\delta _{n+1}^{-1})$
the derivation for $C_{B_{n+1}}(\delta _{n+1}\sigma
_{1}^{-1}d^{-1}(c_{3})d(c_{4})\sigma _{1}\delta _{n+1}^{-1})$ is similar.

For $c_{1},c_{2}\in B_{n}$ define the braids 
\begin{equation*}
d_{1}=\Delta _{n+1}^{2},d_{2}=\sigma _{n}...\sigma
_{2}d^{-1}(c_{1})d(c_{2})\sigma _{2}^{-1}...\sigma _{n}^{-1},d_{3}=\sigma
_{1}...\sigma _{n}^{2}\sigma _{n-1}...\sigma _{1}
\end{equation*}
and 
\begin{equation*}
d_{4}=d_{1}^{-1},d_{5}=d_{2}^{-1},d_{6}=d_{3}^{-1}\text{.}
\end{equation*}

\underline{Proposition 2.5}\medskip

There is a similar proposition in [16]. Let $c_{1},c_{2}\in B_{n}$ and $%
C=C_{B_{n+1}}(\delta _{n+1}\sigma _{1}^{-1}d^{-1}(c_{3})d(c_{4})\sigma
_{1}\delta _{n+1}^{-1})$. The following i) and ii) holds.\newline
i) $d_{1},d_{2},d_{3}\in C.$ \newline
ii) $C^{\prime }=\left\langle d_{1},d_{2},d_{3}\right\rangle $ is an abelian
subgroup of $B_{n+1}$ and hence of polynomial growth. \medskip

\underline{Proof} \medskip

Observe in $B_{n+1}$ 
\begin{equation*}
\sigma _{n}...\sigma _{2}d^{-1}(c_{1})d(c_{2})\sigma _{2}^{-1}...\sigma
_{n}^{-1}=\delta _{n+1}\sigma _{1}^{-1}d^{-1}(c_{1})d(c_{2})\sigma
_{1}\delta _{n+1}^{-1}
\end{equation*}
so $d_{2}\in C$. We know from [16] for arbitrary $p_{i}\in B_{n}$ that
elements of the form $d(p_{i})\sigma _{2}^{-1}...\sigma _{n}^{-1}$ commute
with $d_{3},$ hence $d_{3}$ commutes with $d_{2},$ as $d_{2}$ is of the form 
$d_{2}=(d(p_{1})\sigma _{2}^{-1}...\sigma _{n}^{-1})^{-1}d(p_{2})\sigma
_{2}^{-1}...\sigma _{n}^{-1}$. $d_{1}$ is in the center hence the subgroup $%
C^{\prime }$ is abelian.$\square $

Straight forward variations of our attack are possible. One is could be as
follows. Let $x_{1}^{\prime }=x_{2}^{\prime }$ i.e. solve an MSCSP at (S1)
and the correct this solution using $C_{1}$,$C_{2}$ and we may recover the
actual value of $r$ from one of the correct solutions.

Our ideas can be summarised into the following algorithm.\medskip

\underline{Algorithm 2.1 - Heuristic Algorithm for solving the MSSDP}%
\medskip .

INPUT: The example of the MSSDP given by the equations c 
\begin{eqnarray*}
y_{1} &=&wd(c_{1})\sigma _{1}d(x),\text{ }y_{2}=wd(c_{2})\sigma _{1}d(x) \\
y_{3} &=&wd(c_{3})\sigma _{1}d(x)\text{ and }y_{4}=wd(c_{4})\sigma _{1}d(x)%
\text{;}
\end{eqnarray*}
an objective function $f$ such that $f=0$ when a solution to the MSSDP is
found.

OUTPUT: A solution of the MSSDP.

COMPUTATION:

A. Compute 
\begin{equation}
CE(y_{1},y_{2})=y_{1}^{-1}y_{2}=d^{-1}(x)\sigma
_{1}^{-1}d^{-1}(c_{1})d(c_{2})\sigma _{1}d(x)  \tag{k}
\end{equation}
which is the CSP with $(\sigma _{1}^{-1}d^{-1}(c_{1})w^{-1}wd(c_{2})\sigma
_{1},CE(y_{1},y_{2}))$ with solution $d(x)$%
\begin{equation}
CE(y_{3},y_{4})=y_{3}^{-1}y_{4}=d^{-1}(x)\sigma
_{1}^{-1}d^{-1}(c_{3})d(c_{4})\sigma _{1}d(x)  \tag{l}
\end{equation}
Note we have transformed the MSSDP into an MSCSP involving the shift
operator because equations k and l are an MSCSP in $d(x)$.

B. Using an $XSS$ algorithm e.g. the $USS$ technique compute the solutions $%
s_{1}^{\prime },s_{2}^{\prime }\in B_{n+1}$ 
\begin{eqnarray*}
\delta _{n+1}CE(y_{1},y_{2})\delta _{n+1}^{-1} &=&s_{1}^{\prime }\delta
_{n+1}\sigma _{1}^{-1}d^{-1}(c_{1})d(c_{2})\sigma _{1}\delta
_{n+1}^{-1}s_{1}^{\prime -1} \\
\delta _{n+1}CE(y_{3},y_{4})\delta _{n+1}^{-1} &=&s_{2}^{\prime }\delta
_{n+1}\sigma _{1}^{-1}d^{-1}(c_{3})d(c_{4})\sigma _{1}\delta
_{n+1}^{-1}s_{2}^{\prime -1}\text{.}
\end{eqnarray*}

C. Put $S=(s_{1}^{\prime },s_{2}^{\prime },f(|y_{1}^{-1}\cdot ((y_{1}\cdot
d(s_{1}^{\prime -1})\cdot \sigma _{1}^{-1}\cdot d(c_{1}^{-1}))\ast c_{1}\ast
s_{1}^{\prime })|_{\Delta _{n+1}},|y_{2}^{-1}\cdot ((y_{2}\cdot
d(s_{2}^{\prime -1})\cdot \sigma _{1}^{-1}\cdot d(c_{2}^{-1}))\ast c_{2}\ast
s_{2}^{\prime })|_{\Delta _{n+1}}))=(s_{1}^{\prime },s_{2}^{\prime },f)$ and 
$M=\oslash .$

D. Until a solution is found.

1. Choose a tuple $(t,u,l_{t})$ from $S$ with the smallest $l_{t}$.

2. If $f=0$ then output $t$ (here $t=u$) and 
\begin{equation}
w=y_{1}^{-1}\cdot ((y_{1}\cdot d(s_{1}^{\prime -1})\cdot \sigma
_{1}^{-1}\cdot d(c_{1}^{-1}))\text{.}  \tag{m}
\end{equation}
Note if $f=0$ then by equations j and m we get the actual value of $w$.

3. Otherwise for each $i=1,...,K$ and $j=1,...,L,$ for some natural numbers $%
K$ and $L$.

\ \ \ (i) Compute $t_{i}=t\cdot C_{i},u_{j}=u\cdot C_{j}$ and $f$

\ \ \ (ii). If $(t_{i},u_{j},f)$ belongs neither to $S$ nor to $M$ then to
add it into $S$.

4. Remove the current pair $(t,u,f)$ from $S$ and add it to $M$.\medskip

\underline{3 An Algorithm for the Multiple Simultaneous Shifted
Decomposition Problem Variant MSSDPv}\medskip

\textbf{Definition.} The MSSDPv (multiple simultaneous SDP variant) is as
follows. Let $n\geq 1$ be a fixed integer. For braids $w,x,y_{i},c_{i}\in
B_{\infty }$ $1\leq i\leq n$ find braids $w,x\in B_{\infty }$ such that $%
y_{i}=wd(c_{i})\sigma _{1}d(x)$ where $y_{i}$ are publicly known and $%
w,x,c_{i}$ are secret.\medskip

Clearly the MSSDPv generalises the MSCSPv. The generalisation of the MSSDPv
for the MSSDP is similar to the generalisation of the MSCSPv for the MSCSP.

There are no efficient solutions for solving the MSSDPv one reason for this
is it would mean the SCSP would be easy. In this appendix we propose a
solution for the MSSDPv.

Consider when $n=4$ in the MSSDPv and $c_{1}\neq c_{2}$, $c_{3}\neq c_{4}$
and consider when $n=4$ in the MSSDP and $c_{1}\neq c_{2}$, $c_{3}\neq c_{4}$%
, this is the system of equations 
\begin{eqnarray}
y_{1} &=&wd(c_{1})\sigma _{1}d(x),\text{ }y_{2}=wd(c_{2})\sigma _{1}d(x) 
\TCItag{n} \\
y_{3} &=&wd(c_{3})\sigma _{1}d(x)\text{ and }y_{4}=wd(c_{4})\sigma _{1}d(x) 
\notag
\end{eqnarray}
Again using CE functions we transform the MSSDPv into a shifted MSCSPv type
problem very efficiently.

As the MSSDPv generalises the MSCSPv we can use our algorithm for the MSSDPv
to attack the AAGL protocol in [1].\medskip

\underline{Algorithm 3.1- Heuristic Algorithm for Solving the MSSDPv }%
\medskip

INPUT: The example of the MSSDPv which are equations n above.

OUTPUT: A solution of the MSSDPv.

COMPUTATION:

A. First we compute 
\begin{equation}
CE(y_{1},y_{2})=y_{1}^{-1}y_{2}=d^{-1}(x)\sigma
_{1}^{-1}d^{-1}(c_{1})d(c_{2})\sigma _{1}d(x)  \tag{o}
\end{equation}

which is the CSP with $(\sigma _{1}^{-1}d^{-1}(c_{1})w^{-1}wd(c_{2})\sigma
_{1},CE(y_{1},y_{2}))$ with solution $d(x)$%
\begin{equation}
CE(y_{3},y_{4})=y_{3}^{-1}y_{4}=d^{-1}(x)\sigma
_{1}^{-1}d^{-1}(c_{3})d(c_{4})\sigma _{1}d(x)\text{.}  \tag{p}
\end{equation}

Note we have transformed the MSSDPv into the MSCSPv involving the shift
operator. We observe equations o and p are an MSCSPv in $d(x),\sigma
_{1}^{-1}d^{-1}(c_{1})d(c_{2})\sigma _{1}$, $\sigma
_{1}^{-1}d^{-1}(c_{3})d(c_{4})\sigma _{1}$ and so we then use algorithm 1.1
to recover the ``middle elements'' $\sigma
_{1}^{-1}d^{-1}(c_{1})d(c_{2})\sigma _{1}$ and $\sigma
_{1}^{-1}d^{-1}(c_{3})d(c_{4})\sigma _{1}$: if the ``middle elements'' have
not been recovered then the algorithm has failed and stops here.\ Otherwise
if the algorithm 1.1 fails to find $d(x)$ and hence $x$ we goto step B to
attempt to find $x$ in a different way.

B.Use the algorithm 2.1 to solve the MSSDP to attempt to find $x.$

C. If $x$ has not been found the algorithm has failed otherwise the
algorithm is successful.\medskip

\underline{4 Attack on Dehornoy's Shifted Conjugacy Protocol}\medskip

We can apply our algorithm for the MSSDP to attack the shifted CSP based
protocol [14] in the following specific scenario. We refer the reader to
[14] for details of the DSC protocol.

Alice's authenticates to Bob using $r$ as described in [14]. Then Bob reuses
Alice's $r$ as his random value in the commitment with another user (may be
not Alice) because he assumes it is safe to do so. Following the notation in
[14] of the DSC protocol we can attack that protocol: by letting in the
MSSDP $x=w^{-1}$ , $w=r$, $c_{1}=x_{A},c_{2}=x_{A}^{\prime
},c_{3}=x_{B},c_{4}=x_{B}^{\prime }$; recall from [14] that Alice's
commitment $(x_{A},x_{A}^{\prime })=(r\ast p,r\ast p^{\prime })$ and $%
(p,p^{\prime })$ are publicly known; similarly the notation $%
(x_{B},x_{B}^{\prime })$ refers to Bob commitment. Then when we have found a
value for $r$ using our algorithm for the MSSDP we would expect (because of
equation j) this value to be the actual value of $r$ used in the protocol
instead of a different value satisfying the two equation MSCSP in $r$. Then
when we have this correct value of $r$ we can recover Alice's or Bob's
secret key as follows. $r\ast s$ is publicly known, the attacker waits for $%
b=1$ then computes 
\begin{equation*}
r^{-1}\cdot (r\ast s)\cdot dr\cdot \sigma _{1}^{-1}=ds\text{,}
\end{equation*}
noting we can invert the shift operator on $ds$ we recover $s$ hence
breaking the scheme.\medskip

\underline{5 Comparison of Our Attack with the Longrigg-Ushakov Attack}%
\medskip

We summarize the differences between our new attack and the LU\ attack in
[16].\medskip

i) The LU attack is based on solving the CSP. Our attack is based on the
MSCSP.

ii) The LU attack finds Alice's secret key $s$ or an equivalent key for $s$
in a different way compared to our attack. Our attack, when it is used to
attack the DSC protocol, finds the random braid $r$ in the commitment (and
not a equivalent value for $r$) then using this value of $r$ we recover $s$.

iii) The LU attack is for a general scenario. Our attack is for a more
specific scenario which implies an MSCSP in $r$.

iv) Our simple variation of our attack described above is based on solving
the MSCSP at (S1) and the similar step in the LU attack is based on CSP.
Because the CSP is known to be harder than the MSCSP, hence our attack will
succeed in recovering $s$ when the LU attack fails in some scenarios.

v) The LU attack cannot be used to solve the MSSDP but the LU attack can be
used to solve the SCSP. Our attack solves the MSSDP but does not solve the
version of the SCSP used in the DSC protocol because the $CE$ function does
not exist for $s$.

vi) The LU attack only solves one equation which is the SCSP. Our attack can
be extended in a straightforward way to solve the MSSDPv, MSSDP for any $n$:
to derive the attack for a system of $n$ equations is similar to the
examples for $n=4$ given above. We now give suggestions for selecting secure
parameters.

To defend against the attacks for the AAGL attack we suggest the
following.\medskip

i) Ensure if possible that elements of the centraliser of $u_{i}^{\prime }$
are hard with the CSP ($k$,$zkz^{-1}$).

ii) Ensure that elements of the centraliser of the form $zkz^{-1}$ of $%
u_{i}^{\prime }$ cannot be feasibly computed.

iii) The TTP algorithm may be modified with different choices of $BL,BR$ so
that larger generators are used with the constraint of using RFID
tags.\medskip

To defend against attack for the DSC scheme we suggest the following.

i) The example of the MSSDP in the Dehornoy scheme is hard.

ii) Choose the centraliser $C$ is large.\medskip

\end{document}